\documentclass{aa}  
\usepackage{graphicx}
\usepackage{txfonts}
\usepackage{ulem}
\usepackage{longtable}
\usepackage{verbatim}
\usepackage{url}
\usepackage{float,color}
\usepackage{subcaption}
\usepackage{xcolor}

\usepackage{natbib}
\usepackage{hyperref}
\hypersetup{
    colorlinks=true,
    linkcolor=blue,
    filecolor=magenta,      
    urlcolor=cyan,
    citecolor=blue
}

\bibpunct{(}{)}{;}{a}{}{,}

\newcommand{\msun}{\ensuremath{\rm M_\odot}}
\newcommand{\feh}{\ensuremath{\rm{[Fe/H]}}}
\newcommand{\teff}{\ensuremath{\rm T_{\rm eff}}}
\newcommand{\logg}{\ensuremath{\rm \log{g}}}

\newcommand{\Rgc }{\ensuremath{{\rm R_{\rm gc}}}}
\newcommand{\afe}{\ensuremath{\rm [\alpha/Fe]}}

\newcommand{\kms}{\ensuremath{\rm km s^{-1}}}
\newcommand{\vphi}{\ensuremath{\rm V_\phi}}
\newcommand{\vr}{\ensuremath{\rm V_r}}
\newcommand{\vz}{\ensuremath{\rm V_z}}

\newcommand{\jr}{$J_r$}
\newcommand{\lz}{$L_z$}
\begin{document} 


\title{The Prince and the Pauper: Evidence for the early high-redshift formation of the Galactic $\alpha$-poor disc population \thanks{Based on observations made with the ESO/VLT, at Paranal Observatory, under program 188.B-3002 (The Gaia-ESO Public Spectroscopic Survey, PIs G. Gilmore and S. Randich). Also based on observations under programs 171.0237 and 073.0234.}}
   
\author{Matthew Raymond Gent \inst{1,2} \and
        Philipp Eitner \inst{1,2} \and
        Aldo Serenelli \inst{5,6} \and
        Jennifer K. S. Friske \inst{3} \and
        Sergey E. Koposov \inst{7,8} \and
        Chervin F. P. Laporte   \inst{4} \and
        Tobias Buck \inst{10,11} \and 
        Maria Bergemann \inst{1,9}}
          
\institute{
Max-Planck Institute for Astronomy, 69117 Heidelberg, Germany
\label{inst1}
\and
Ruprecht-Karls-Universit\"at, Grabengasse 1, 69117 Heidelberg, Germany
\label{inst2}
\and 
 Mullard Space Science Laboratory, University College London, Holmbury St. Mary, Dorking, Surrey, RH5 6NT, UK
\label{inst3}
\and
Institut de Ciencies del Cosmos (ICCUB), Universitat de Barcelona (IEEC-UB), Martii i Franques, 1, E-08028 Barcelona, Spain
\label{inst4}
\and
Institute of Space Sciences (ICE, CSIC), Carrer de Can Magrans S/N, E-08193, Cerdanyola del Valles, Spain
\label{inst5}
\and
Institut Estudis Espacials de Catalunya (IEEC), Carrer Gran Capita 2, E-08034, Barcelona, Spain
\label{inst6}
\and
Institute for Astronomy, University of Edinburgh, Royal Observatory, Blackford Hill, Edinburgh EH9 3HJ, UK
\label{inst7}
\and Institute of Astronomy, University of Cambridge, Madingley Road, Cambridge CB3 0HA, UK
\label{inst8}
\and
Niels Bohr International Academy, Niels Bohr Institute, University of Copenhagen, Blegdamsvej 17, DK-2100 Copenhagen, Denmark
\label{inst9}
\and
Universit\"{a}t Heidelberg, Interdisziplin\"{a}res Zentrum f\"{u}r Wissenschaftliches Rechnen, Im Neuenheimer Feld 205, D-69120 Heidelberg, Germany
\label{inst10}
\and
Universit\"{a}t Heidelberg, Zentrum f\"{u}r Astronomie, Institut f\"{u}r Theoretische Astrophysik, Albert-Ueberle-Stra{\ss}e 2, D-69120 Heidelberg, Germany
\label{inst11}
}

\date{}

 
\abstract
%
{The presence of [$\alpha$/Fe]-[Fe/H] bi-modality in the Milky Way disc has animated the Galactic archaeology community since more than two decades.}
{Our goal is to investigate  the chemical, temporal, and kinematical structure of the Galactic discs using abundances, kinematics, and ages derived self-consistently with the new Bayesian framework SAPP.}
{We employ the public Gaia-ESO spectra, as well as \textit{Gaia} EDR3 astrometry and photometry. Stellar parameters and chemical abundances are determined for $13\,426$ stars using NLTE models of synthetic spectra. Ages are derived for a sub-sample of $2\,898$ stars, including subgiants and main-sequence stars. The sample probes a large range of Galactocentric radii, $\sim$ 3 to 12 kpc, and extends out of the disc plane to $\pm 2$ kpc.}
{Our new data confirm the known bi-modality in the [Fe/H] - [$\alpha/$Fe] space, which is often viewed as the manifestation of the chemical thin and thick discs. The over-densities significantly overlap in metallicity, age, and  kinematics, and none of these is a sufficient criterion for distinguishing between the two disc populations. Different from previous studies, we find that the $\alpha$-poor disc population has a very extended [Fe/H] distribution and contains $\sim 20 \%$ old stars with ages of up to $\sim 11$ Gyr.}
{Our results suggest that the Galactic thin disc was in place early, at look-back times corresponding to redshifts $z \sim 2$ or more. At ages $\sim$ 9 to 11 Gyr, the two disc structures shared a period of co-evolution. Our data can be understood within the clumpy disc formation scenario that does not require a pre-existing thick disc to initiate a formation of the thin disc. We anticipate that a similar evolution can be realised in cosmological simulations of galaxy formation.}


\authorrunning{Gent et al.}
\titlerunning{Evidence for the early high-redshift
formation of the Galactic $\alpha$-poor disc population}

\maketitle
%
%
%
\section{Introduction} \label{sec:introduction}

Still, the structure and evolution of the Galactic disc remains one of the most complex problems in studies of Galaxy formation. Since the discovery of the thick disc \citep{Gilmore1983}, much work focused on the bi-modality in the space of chemical abundances \citep[e.g.][]{Bensby2005, Reddy2006, Bovy2012, Recio-Blanco2014, Duong2018}. Many studies based on stars in the solar neighbourhood and beyond  pointed out the existence of two populations, $\afe$-rich and $\afe$-poor, partly overlapping in metallicity \citep[e.g.][]{Fuhrmann1998, Nidever2014,guiglion2023gaia}, age \citep[e.g.][]{Bensby2014, Feuillet2019}, and kinematics \citep[e.g.][]{Ruchti2011, Lee2011, Kordopatis2011, Hayden2015}. These stellar populations have been deemed as the "thin" and the "thick" disc, and their morphological parameters have since then been subject of a major interest \citep{Bland-Hawthorn2016}. It has been shown different physical processes may influence the formation and evolution of sub-structure in the disc, including multiple infall \citep[e.g.][]{Chiappini1997,Spitoni2019}, radial migration \citep[e.g.][]{Schoenrich2009a, Schoenrich2009b, Loebman2011, Minchev2013} and radial mixing caused by satellites \citep{Quillen2009}, growth induced by mergers \citep[e.g.][]{Villalobos2008, Read2008, Villalobos2010}, gas-rich accretion and mergers \citep[e.g.][]{Brook2004, Stewart2009, Grand2018, Buck2020}, local gas instabilities in discs associated with turbulence \citep[e.g.][]{Noguchi1999,Bournaud2009,Clarke2019}, dynamical interaction with cold dark matter sub-halos \citep[][]{Hayashi2006, Kazantzidis2009}, galactic winds \citep[][]{Moster2012}, and early outflows \citep[][]{Khoperskov2021}. It has also been demonstrated that the chemical bi-modality is a relatively rare phenomenon in L$^*$ galaxy discs \citep[][]{Mackereth2018, Gebek2021}, but see \citet{Khoperskov2021}. More recent studies address in more detail the spatial variability of the bi-modality \citep{Hayden2015, Bovy2016, Nandakumar2020}, finding that \afe-rich component is more centrally concentrated, whereas the \afe-poor component has a larger radial extent \citep{Haywood2019,Sahlholdt2022}.

Therefore, the question of whether the populations are indeed distinct stellar components with a separate formation history still remains open. 

In this work, we perform of a detailed analysis of the chemical, temporal, and kinematical distribution functions of the Galactic disc populations, using data from the Gaia-ESO survey and \textit{Gaia}. The key difference between the Gaia-ESO and other comparable medium- and high-resolution stellar surveys, RAVE, APOGEE, and GALAH, is its photometric coverage. Most stars in the Gaia-ESO selection are rather faint ($14 \lesssim$ G$_{\rm  mag}$ $\lesssim 18$ mag), and as outlined in \citep{Gilmore2012, Randich2013,Stonkute2016} this selection allows to uniquely address the parameter space of the inner and outer thin and thick disc, and the thick disc to halo transition, beyond what is possible with brighter and more local samples.

The paper is organised as follows, in Sect. \ref{sec:observations} we discuss  the observational sample. Sect. \ref{sec:stellarparameters} outlines the approach used for the determination of stellar parameters and chemical abundances. In Sect. \ref{sec:ages}, we briefly state how the ages are determined. Sect. \ref{sec:validation} deals with the validation of stellar parameters and ages. The results are presented in Sect. \ref{sec:results}, where we focus on the distributions of chemical abundances, combined with kinematics and ages. Further, we discuss the results in the context of previous observational and theoretical findings, and we close the paper with conclusions in Sect. \ref{sec:conclusion}.
\section{Observations} \label{sec:observations}
In this work,  we rely on targets observed within the Gaia-ESO large spectroscopic survey \citep{Gilmore2012,Randich2013}. In the latest public data release (DR4), spectra for over 10$^5$ are available, and we use all spectra taken with the HR10 setting of Giraffe spectrograph\footnote{The NLTE grids employed in this work \citep{Kovalev2019,Gent2022a} cover the corresponding wavelength regime}. The HR10 data are available for $55\,761$ stars. The signal-to-noise (SNR) distribution of the sample is very broad and ranges from $2$ to a few $100$ per pixel. 
\begin{figure}[ht!]
\hbox{
\includegraphics[width=0.5\columnwidth]{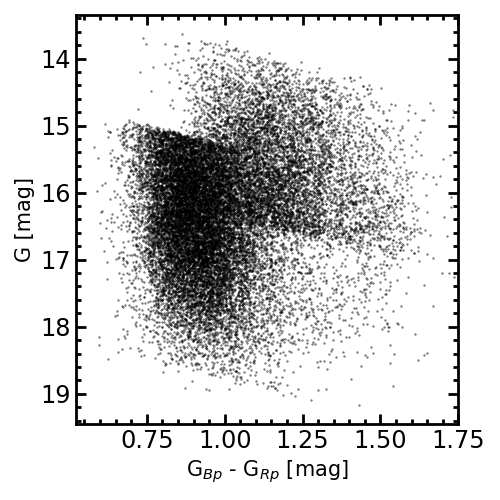}
\includegraphics[width=0.5\columnwidth]{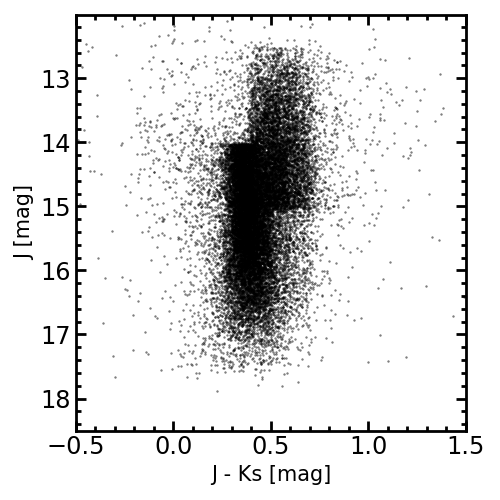}}
\caption{Photometry of the observed sample. Left panel shows the distribution in the \textit{Gaia} magnitudes, G vs G$_{Bp}$ - G$_{Rp}$. Right panel shows the distribution in the VHS magnitudes, J vs J - K$_\text{s}$ (see Sect. 2).}
\label{fig:phot_CMD}
\end{figure}

Fig. \ref{fig:phot_CMD} shows the targets in the photometric colour-magnitude (CMD), J versus J - K$_{\rm s}$, plane, where J and K$_{\rm s}$ are VISTA photometric filters \citep{McMahon2013}. The apparent regular distribution is caused by the photometric selection of targets in the input Gaia-ESO catalogue. For the Giraffe catalogue, the following basic selection scheme was used: $0.00 \leq (\text{J} - \text{K}_{\rm s}) \leq 0.45$ and $14.0 \leq J \leq 17.5$ for the blue box, and  $0.40 \leq (\text{J} - \text{K}_{\rm s}) \leq 0.70$ and $12.5 \leq \text{J} \leq 15.0$ for the red box. The boxes were defined to maximise the probability of observing targets in all Galactic components, the discs and the halo, therefore the target densities vary drastically, and to account for this effect, the boxes were slightly extended in order to optimise the fiber occupancy in each field. In particular, in the fields, where number of targets exceeded the number of fibers - as in low latitude fields -, additional selection criteria were used, such as shifting the boxes by the mean value of extinction in a given field $0.5$ E(B - V). This procedure leads to a characteristic spread of the distribution along the x-axis, as seen in Fig. \ref{fig:phot_CMD}. The relative distribution is such that the majority of targets (80 $\%$) are in the blue box, whereas stars in the red box account for about 20 $\%$ of the sample. The blue box targets include main-sequence, turnoff, and subgiant stars, and the red box was optimised for red clump stars, however because of the extension of the boxes a certain overlap exists. For the detailed description of the input catalogue, we refer to \citet{Stonkute2016}. We note that this selection implies that most targets in the Gaia-ESO HR10 sample are much fainter, $14 \lesssim$ G$_{\rm  mag}$ $\lesssim 18$ mag, compared to other surveys such as RAVE, APOGEE, or GALAH. The selection also has a strong effect on the  properties of our sample, in the plane of astrophysical parameters. Specifically \citep[e.g.][]{Bergemann2014,Thompson2018}, the distribution of our sample is preferentially skewed towards older populations with slightly sub-solar metallicities and we explore this issue in more detail in  Sect. \ref{sec:selection}.

\begin{figure}[ht!]
\centering
\includegraphics[width=\columnwidth]{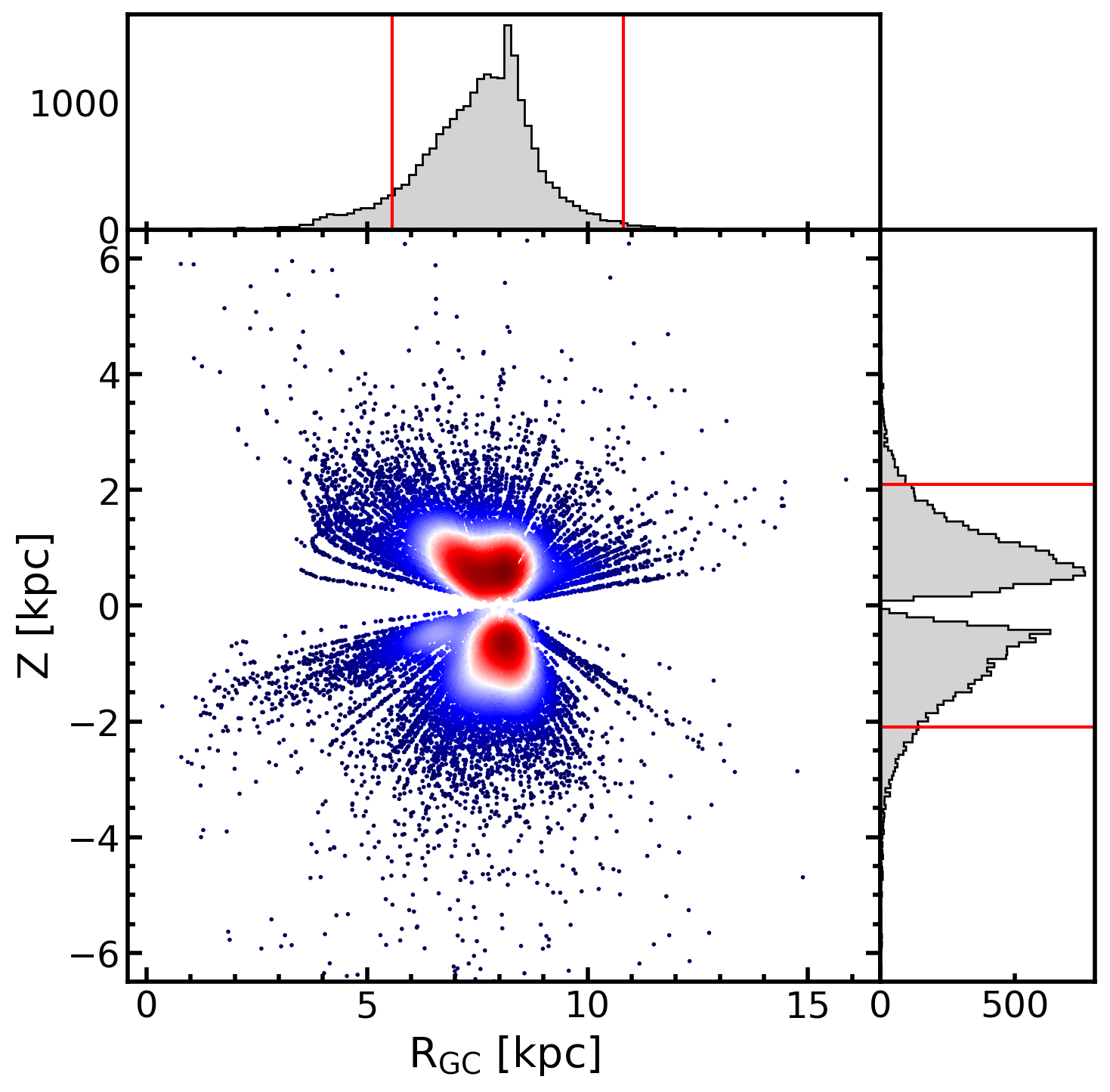}
\caption{Spatial distribution of the observed sample. The vertical axis represents the height above the disc plane in kpc and the horizontal represents the Galactocentric radius in kpc. The colour scale shows normalised density from 0.07$\%$ (dark blue) to 100$\%$ (dark red).}
\label{fig:space_distr}
\end{figure}

We complement these data with the proper motions, photometry, and extinction from the EDR3 \textit{Gaia} catalogue \citep{Gaia-Collaboration-2020}. The cross-match between every Gaia-ESO spectrum and \textit{Gaia} ED3 catalogue was performed on grounds of angular position within a $1.0$ arcsec tolerance (cone search). Distances and their uncertainties were adopted from  \cite{Bailer-Jones2021}. The spatial distribution of the sample is shown in Fig. \ref{fig:space_distr}. Most of these objects are confined closer to the plane with altitudes of up to $~2$ kpc and they probe a range of Galactocentric radii from $\sim 5$ to $12$ kpc. The 3D space velocities\footnote{In this work, we use the Galactocentric cylindrical coordinate system. So that V$_{\rm r}$, V$_{\phi}$, V$_z$ are the components of the full 3D space velocity pointing  towards Galactic center R, in the direction of rotation $\phi$, and vertically relative to the disc mid-plane, respectively.} for the sample are calculated using the Python package 'astropy' \citep{astropy2013,astropy2018}. The accuracy of the astrometric information is high enough to yield the velocities with an uncertainty of $\lesssim 5$ kms$^{-1}$.

\section{Methods} 
\subsection{Stellar parameters} \label{sec:stellarparameters}
The homogeneity, accuracy, and precision of stellar parameters is essential given by the scientific goals of this study. However, our analysis of the Gaia-ESO sixth internal data release (iDR6)\footnote{\url{https://www.gaia-eso.eu/}} \citep{Gilmore2012,Randich2013} show artificial ridges and bifurcations in the space of stellar parameters and their uncertainties. It suffers from some loss of precision owing to the complex homogenisation and cross-calibration procedure employed.Therefore, this does not allow for a robust analysis of the distribution functions in the space of astrophysical parameters and ages. 

\begin{figure}[ht]
\includegraphics[width=\columnwidth]{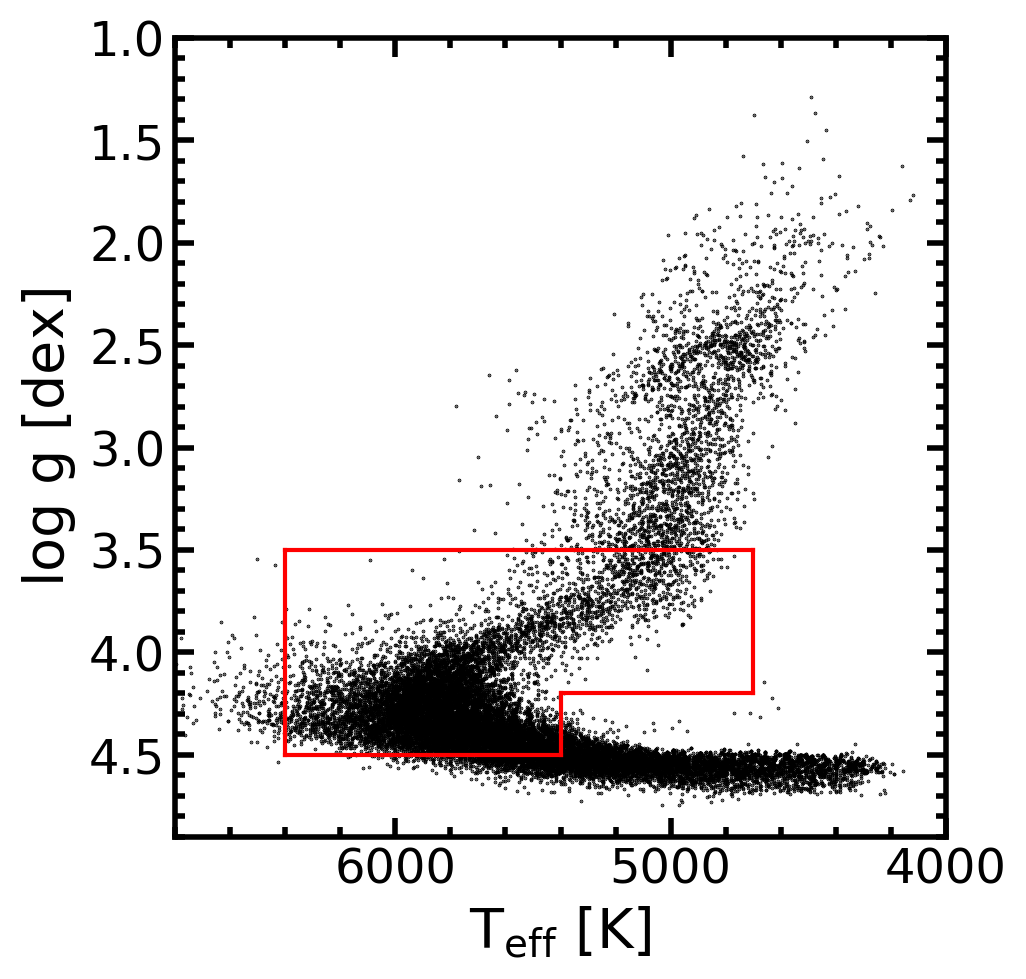}
\caption{The distribution of the observed sample in the \teff-\logg~plane. The targets enclosed within the red box represent the sample used for the analysis of ages.} 
\label{fig:parameters}
\end{figure}
We have therefore opted to re-analyse the spectra using the Bayesian SAPP pipeline described in \citet{Gent2022a}. This method was shown to provide atmospheric parameters, including $\teff$, $\logg$, $\feh$, and abundances (Mg, Ti, Mn), across a broad range of stellar parameters: $4000 \lesssim \teff \lesssim 7000$ K, $1.2 \lesssim \logg \lesssim 4.6$ dex, and $-2.5 \lesssim \feh \lesssim +0.6$ dex. Extensive tests of the NLTE synthetic spectral grids and the Payne model \citep{Ting2019} based on these can be found in \citet{Kovalev2019}. The latter study also presented a detailed spectroscopic analysis of the benchmark stars, including main-sequence stars, subgiants, and red giants, and $742$ stars probing the entire evolutionary sequence in $13$ open and globular clusters. In \cite{Gent2022a}, the code was developed to carry out the full Bayesian analysis, by combining the probabilities obtained from the spectroscopy, photometry, astrometry, and asteroseismic data analysis modules. 

The results of the SAPP analysis are shown in Fig. \ref{fig:parameters}, where the entire Gaia-ESO HR10 sample with SNR $> 20$ is plotted in the $\teff$-$\logg$ plane. In total, we have $13\,426$ stars with reliable stellar parameters The characteristic uncertainties of stellar parameters are of the order $51$ K for $\teff$, $0.04$ dex for $\logg$, $0.05$ dex for [Fe/H], $0.06$ dex for [Mg/Fe]. These uncertainties represent the combined estimates derived from the shape of the multi-dimensional posterior PDF, and thus, they account for the statistical uncertainties (those of the observed data) and for the systematic (differences between the individual PDFs derived from the photometric, astrometric, and spectroscopic data). For more details on the error analysis, we refer the reader to \cite{Gent2022a}. 
\subsection{Stellar ages} \label{sec:ages}
One important component of this study is the availability of ages. Ages are derived using the Bayesian pipeline BeSPP presented in \citet{Serenelli2013}, which was also applied to the analysis of the first Gaia-ESO data release in \citet{Bergemann2014}. The code relies on the GARSTEC grid of stellar evolutionary tracks \citep{Weiss2008} (also used in SAPP) that covers the mass range from 0.6 to 5.0 $\msun$ with a step of 0.02 $\msun$ and metallicity from $-2.50$ to $+0.60$ with a step of 0.05 dex. The spectroscopically derived [$\alpha$/Fe] ratio is taken into account for the age determinations by modifying the derived [Fe/H] following the prescription [Fe/H] $\longrightarrow$ [Fe/H] + 0.625 [$\alpha$/Fe], which agrees with the traditionally used modification by \cite{Salaris1993} within 10\% in the range -0.45 $<$[$\alpha$/Fe]$<$+0.45.


The analysis of ages is limited to subgiants and upper main-sequence stars (including turn-off, TO), because of the strong degeneracy typically identified between tracks of different ages and metallicities for the lower MS and RGB phases. This selection is made by limiting the effective temperature and surface gravity to: 4700 - 6700 K, 3.5 - 4.5 dex for subgiants and turn-off stars, and 5400 - 6700 K, 3.5 - 4.2 dex for the upper main-sequence. We also include the stars with accurate abundances and ages analysed in \cite{Bergemann2014}. These stars are part of the Gaia-ESO sample. We further limit our sample with a maximum of $0.1$ dex in [Fe/H] error. The uncertainties of stellar parameters are small enough to ensure that our selection retains most of the subgiants in the sample and it minimises the contamination by lower main-sequence stars. 

In the next section, we describe the validation of our stellar parameters and ages using different inputs and methodological approaches. We then present the final sample with accurate ages that is used for the analysis of the properties of the Galactic disc.
\subsection{Validation of parameters}\label{sec:validation}
The quality of stellar parameters is important within the scope of this paper. In Paper 1 \citep{Gent2022a}, we presented a careful validation of our stellar analysis pipeline and its outputs (including metallicities, masses, and ages), using a sample of benchmark stars with independently determined stellar parameters, including interferometic $\teff$ and ages constrained by asteroseismology. We showed that a combination of spectroscopy, astrometry, and photometry in the Bayesian framework yields metallicities accurate to $0.02$ dex and ages with the precision of $\sim 0.6$ and accuracy of $\sim 1$ Gyr, in line with results of similar earlier studies \citep[e.g.][]{Serenelli2013, Schoenrich2014}. Whereas the focus of our work in Paper 1 was on main-sequence stars and subgiants, and the same type of observational information, that is the Gaia-ESO spectra, \textit{Gaia} photometry and astrometry, was employed, the difference with respect to present study is the use of global asteroseismic quantities that are not available for the majority of stars in the present sample.

\begin{table}
\begin{center}
\caption{Sensitivity of the astrophysical parameters of the stellar sample to changes in the input data and in methodology. See text.}
\label{tab:partest}
\renewcommand{\footnoterule}{} 
\setlength{\tabcolsep}{3pt}
\begin{tabular}{l | ccccc}
\hline\hline       
  & $\Delta$ \rm{[Fe/H]} & $\Delta$ $\teff$ & $\Delta$ $\logg$ & Age \\
  &           dex &            K &  dex        & Gyr \\
\hline
  IRFM $\teff$ &  0.01 $\pm$ 0.04 & 16 $\pm$ 52  & -0.06 $\pm$ 0.14 & - \\
  d $\pm$ 10\% &  0.00 $\pm$ 0.02 & 2 $\pm$ 12 & 0.00$ \pm$ 0.08 & - \\
   $J$, $K_s$   & - & - & - & 0.9 $\pm$ 2.2 \\
  BeSPP - SAPP & - & - & - & 0.5 $\pm$ 0.4 \\
\hline               
\end{tabular}
\end{center}
\end{table}
Therefore, here we present additional tests in order to investigate the properties of data errors in the parameter space that is relevant to our conclusions. First, we carry out the analysis of atmospheric parameters of stars using the infra-red flux (IRFM) method \citep{Casagrande2010, Casagrande2021}. The results obtained by applying the IRFM technique to our main sample are provided in Table \ref{tab:partest}. The effective temperatures are in agreement with the reference SAPP values to 16 $\pm$ 52 K, whereas the resulting surface gravities and metallicities change by $-0.06 \pm 0.14$ dex and $0.01 \pm 0.04$ dex, respectively, if $\teff$ is derived from IRFM instead of spectroscopy. We also investigate the quality of surface gravities calculated by the SAPP by applying a systematic shift to distances that are adopted from \citep{Bailer-Jones2021}. The shift was estimated through the analysis of typical uncertainties of distances: the average distance error in our sample is of the order 8 $\%$ pc with the majority of stars having the error of $\sim 5\%$. The resulting stellar parameters calculated with the offset distances are also provided in Table \ref{tab:partest}. The shift has no significant effect on Teff, with the average difference of $2$ K and a scatter $12$ K, whereas surface gravities and metallicities are affected at the level of $0.00 \pm 0.08$ dex and $0.00 \pm 0.02$ dex, respectively.

In the third step, we explore the sensitivity of ages to the methodology by performing the analysis of ages using an alternative approach presented in \citet{xiang2022}, where only $K_s$ magnitudes are used circumventing the need for surface gravities. In the latter case, we make use of either only $K_s$, or $K_s$ and $J$ magnitudes from 2MASS and the synthetic photometry. Comparing the resulting ages with the ages calculated self-consistently within the Bayesian framework, we find that the effect on ages is maximum at slightly sub-solar metallicities, with the bias and scatter of $\sim~0.9$ Gyr and $\sim~2.2$ Gyr, respectively. Interestingly, more metal-poor stars, [Fe/H] $\lesssim -0.7$ dex, are least affected by the choice of the method, with the age bias of only $0.3$ Gyr and scatter of $1.4$ Gyr.
Finally, we compare the ages obtained using BeSPP with the age estimated by the SAPP, following \citet{Gent2022a}. Since the codes assume a similar algorithm, the same input data and evolutionary tracks, this comparison primarily demonstrates the internal precision of ages. We find the both codes are in excellent agreement, with a $0.5$ Gyr bias and a scatter of $0.4$ Gyr.

Based on these tests, we furthermore select only those stars, for which ages consistent to better than $1$ Gyr are obtained with both methods and with both codes, that yields a final sample of $2\,898$ stars with the characteristic uncertainty of ages ranging from $0.7$ to $2$ Gyr and the mean error of $1.3$ Gyr.
\subsection{Survey selection function}\label{sec:selection}
In order to assess the influence of the survey selection function on our data set, we followed the methodology of \citet{Bergemann2014} and \citet{Thompson2018}. To ensure self-consistency in the analysis, the same evolutionary tracks were used (Sect. \ref{sec:ages}). Firstly, a complete population of stars has been generated assuming the Salpeter initial mass function, a constant star formation rate, and a uniform and age-independent metallicity [Fe/H] distribution. This [Fe/H] population exhibits a trend, which reflects simply the IMF and the stellar evolution lifetime, that is shorter at lower [Fe/H] and same mass. We then remove stars outside the $\teff$-$\logg$ box used in our analysis (Sect.~\ref{fig:parameters}). In the second step, this Mock dataset is used to determine the completeness fraction by calculating the ratio of photometrically selected targets relative to the complete  sample. This is done separately for the blue and red photometric boxes. 

\begin{figure}[ht!]
\centering
\includegraphics[width=1\columnwidth]{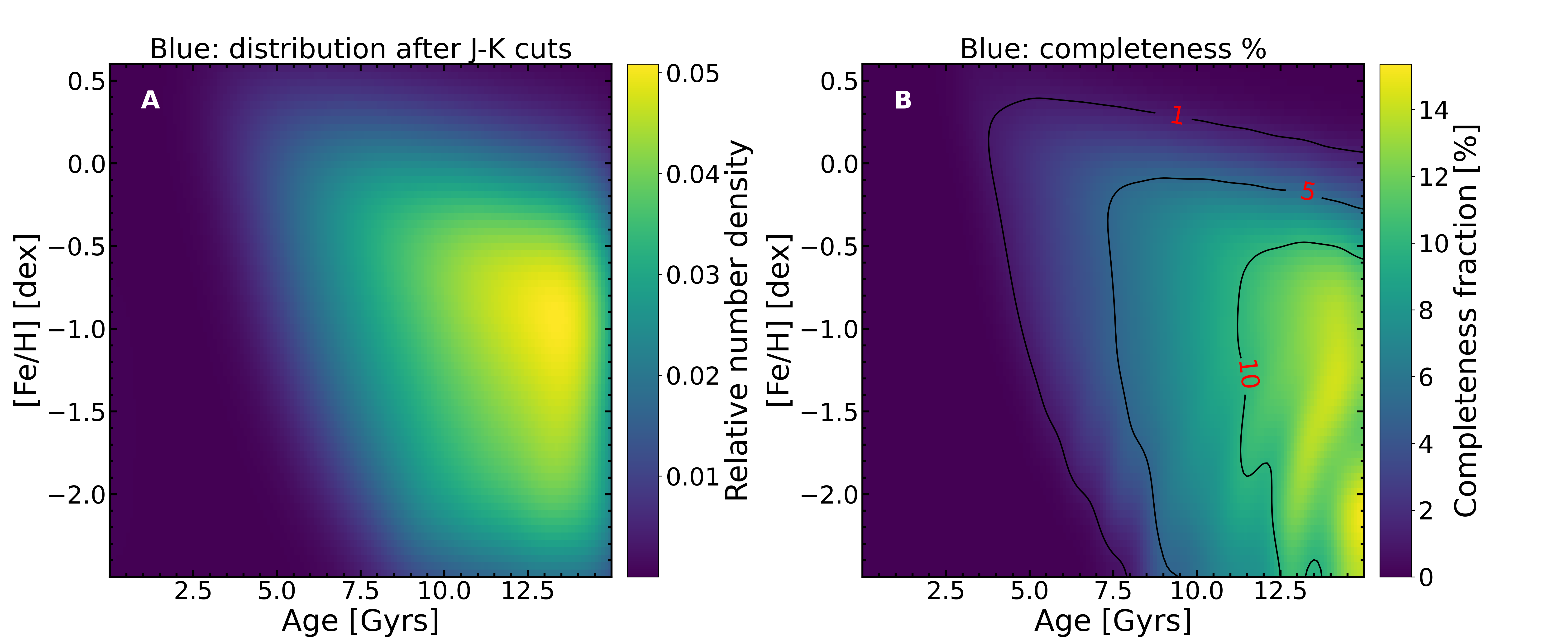}
\includegraphics[width=1\columnwidth]{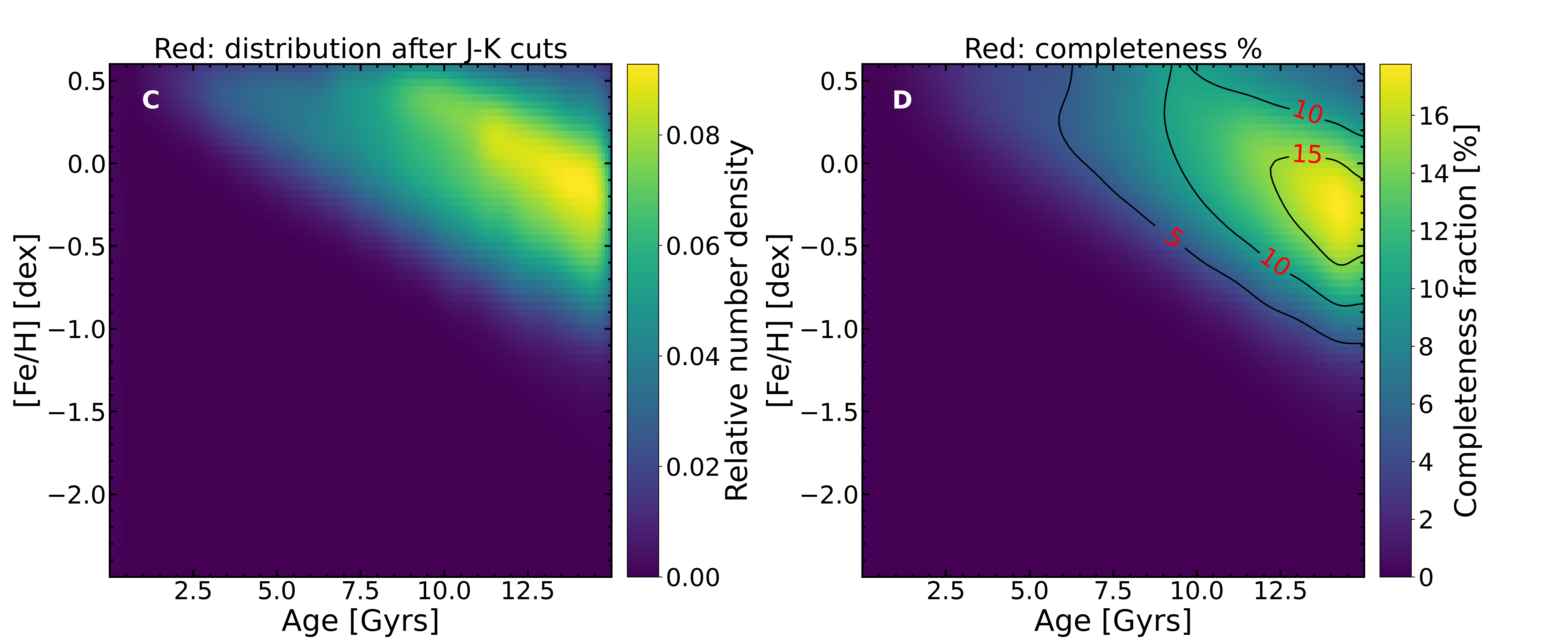}
\caption{Synthetic stellar population to simulate the selection effect of Gaia-ESO survey and of the observed stellar sample. Here the results of the simulation at $1$ kpc are shown for the blue box (top row) and red box (bottom row) as defined in Sect. \ref{sec:observations}. Right: the completeness fraction, defined as the ratio of stars with cut to the number of stars without cut, i.e. the larger the number the more stars are retained in the population after applying the colour cuts (Sect. \ref{sec:observations}). In case of no bias, the completeness fraction is unity $1$.}
\label{fig:distance_completeness}
\end{figure}

Figure \ref{fig:distance_completeness} shows for a given distance of $1$ kpc, the relative stellar density (left) and completeness (right) of the Mock stellar population from the red and blue selection boxes. The distance was chosen as representative of the bulk fraction of stars in the observed Gaia-ESO sample, but careful inspection of the simulated fractions suggests that the distribution is qualitatively similar at distances at $0.5$ kpc or $2$ kpc. It can be concluded that the distribution of stars in the age-metallicity plane suffers from a systematic bias, which is primarily caused by the colour cuts adopted in the Gaia-ESO survey. These cuts lead to a very pronounced depletion in the fraction of young stars with ages $\lesssim~7$ Gyr, although the effect slightly depends on metallicity. The red box additionally skews the distribution towards old metal-rich stars, whereas the blue box is primarily sensitive to old metal-poor stars. This situation is rather similar to the completeness for the Gaia-ESO UVES sample described by  \citet{Bergemann2014} and \citet{Thompson2018}. 
\section{Results} \label{sec:results}

\subsection{Chemical abundances}
\label{sec:chem_abund}

\begin{figure}[ht]
\centering
\includegraphics[width=1\columnwidth]{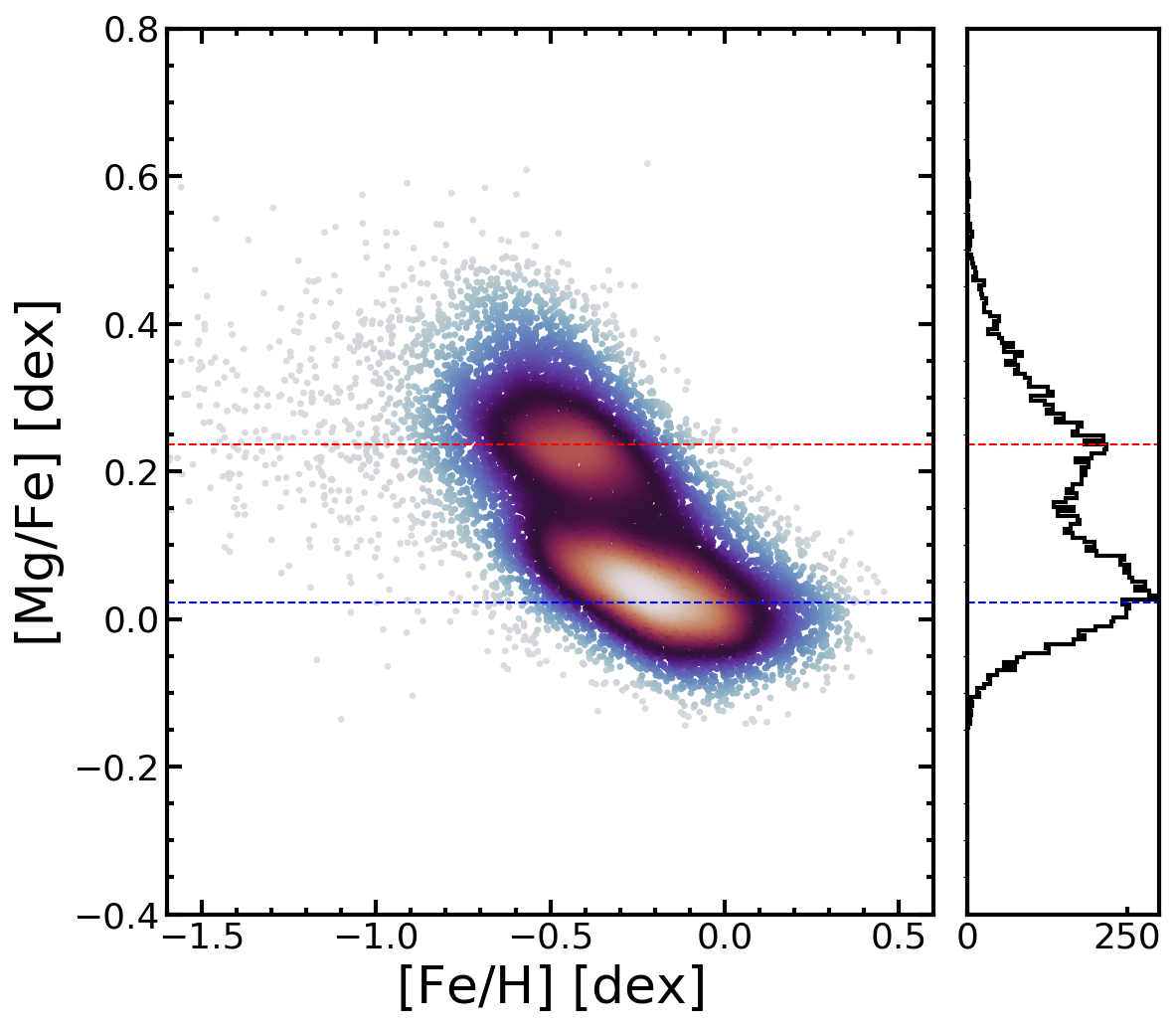}
\caption{The densities of NLTE abundances of  [Mg/Fe] as a function of [Fe/H]. The dashed lines represent the average [Mg/Fe] value for stars which are $\alpha$-rich (red) and $\alpha$-poor (blue).}
\label{fig:mgfe_v_feh_whole}
\end{figure}

The [Mg/Fe] abundance ratios of our sample against metallicity are shown in Fig.\ref{fig:mgfe_v_feh_whole}. The number of stars with reliable abundances is the same sample as Sect. \ref{sec:stellarparameters}. is Here we limit the analysis of [$\alpha$/Fe] to the abundance of Mg, because no particular sub-structure is visible in the distribution of Ti or Mn abundance ratios. We also remind the reader that Mg is a key $\alpha$ element produced by $\alpha$ capture reactions during the hydrostatic C and Ne burning phases in massive stars \citep{Clayton1968, Woosley1995}. Therefore, it is reasonable to adopt the notation of Mg as a representative of the $\alpha$-group.

As previous studies showed, we also see a prominent bi-modality in the [Mg/Fe] abundance ratios, which is manifested as two over-densities separated at [Mg/Fe] $\approx +0.15$ dex across the entire metallicity range, $-1.5 \approx \feh \approx -0.2$ dex. The low-$\alpha$ component peaks at [Mg/Fe] $\approx +0.05$ dex and the high-$\alpha$ component at [Mg/Fe] $\approx +0.24$ dex. It shall be stressed that no component of the analysis, neither the observed data (spatial distribution) nor the grid (models), have any known non-linear dependence that could lead to this discontinuity in the space of Mg and Fe abundances. In particular, in the spectroscopic grid used in the chemical abundance analysis, all elemental abundances have a random uniform distribution. In agreement with the visible over-densities, we choose to assign stars to the $\alpha$-rich population, if their associated [Mg/Fe] abundance is above $0.13$ dex, and to the $\alpha$-poor disc otherwise. Throughout the text, we will proceed to call these two sets of stars "$\alpha$-rich" and "$\alpha$-poor" populations.

Comparing our distributions with literature, we find an overall satisfactory agreement, although it should be noted that owing to a combination of factors, including vastly different spatial-photometric coverage of observational surveys, their observational strategy, and incomplete volume sampling, certain differences arise that render a one-to-one comparison of chemical abundances in a given volume of the Galaxy impossible. Nonetheless, it appears that our distributions are consistent with the distributions seen in the previous DRs of the Gaia-ESO \citep[e.g.][]{Mikolaitis2014, Recio-Blanco2014, Kordopatis2015, Guiglion2015}, as well as independent studies \citep[e.g.][]{Adibekyan2013, Bensby2014}. In the latter work, a prominent separation is detected in [Ti/Fe] abundance ratios, whereas the [Mg/Fe] ratios show a more continuous distribution. This could be the consequence of the spatial coverage of the sample. The study by Bensby et al. focuses on the nearby stars in the solar neighbourhood, whereas our sample probes more extended regions in the $\Rgc$ - z space, reaching $3 < \Rgc < 13 $ kpc and $|z| \approx \pm 3$ kpc. A similar distribution of the low and high-$\alpha$/Fe populations is seen in the APOGEE sample  \citep[e.g.][]{Hayden2015}. One should note, however, that their $\alpha$ parameter refers to the average of different elements\footnote{O, Mg, Si, S, Ca, and Ti}, and so their distributions are not directly comparable to the present sample. The GALAH survey results \citep{Bland-Hawthorn2019, Buder2021} are also consistent with our distributions. In the chemical distributions of the RAVE stellar sample \citep{Steinmetz2020}, the $\alpha$-rich component is discernible at [$\alpha$/Fe] $\approx +0.45$, that is somewhat higher compared to our data and the samples from APOGEE\footnote{The Ti abundances from the APOGEE SDSS-DR16 appear to be unreliable.}\citep{Hayden2015, Jonsson2020}, although consistent within the uncertainties of both samples. 

\subsection{Kinematics and abundances}\label{subsec:kinematics}

For our kinematic analysis, we use a right-handed Galactocentric cylindrical coordinate system, where the velocity component $\vr$ points towards the Galactic centre, $\vphi$ in opposition direction of the rotation and $\vz$ vertically down out of the disk. By converting RA, DEC and proper motions to Galactocentric coordinate frames executed by python package "astropy" \citep{astropy2013,astropy2018}, we derive Cartesian Galactocentric coordinates and velocities. Thus,
\begin{equation}
\label{eqn:vr}
\vr = (XV_X + YV_Y)/R\\ 
\end{equation}
\begin{equation}
\vphi = -(YV_X - XV_Y)/R
\end{equation}
where R = $\sqrt{X^2+Y^2}$. We use Solar Galactocentric coordinates r$_\odot$ = (X$_\odot$,Y$_\odot$,Z$_\odot$)$^T$ = (-8.122,0,0)$^T$ kpc and v$_\odot$ = (U$_\odot$,V$_\odot$,W$_\odot$)$^T$ = (11.1,12.24,7.24)$^T$ km/s \citep{Schoenrich2010} thus defining the Solar Galactocentric radius as R = 8.122 kpc \citep{Drimmel2018}. Then V$_X$ = U, and V$_Y$ = V + V$_c,\odot$, where V$_c,\odot$ is the circular velocity value at Solar radius = 233.4 km/s \citep{Drimmel2018}\footnote{Note the minus sign in otherwise standard definition of $\vphi$, we use this to make clockwise solar circular motion positive. We later treat specific angular momentum \lz\ the same in Sect. \ref{sec:orbits}.}.

\begin{figure*}[ht!]
\centering
\includegraphics[width=0.7\textwidth]{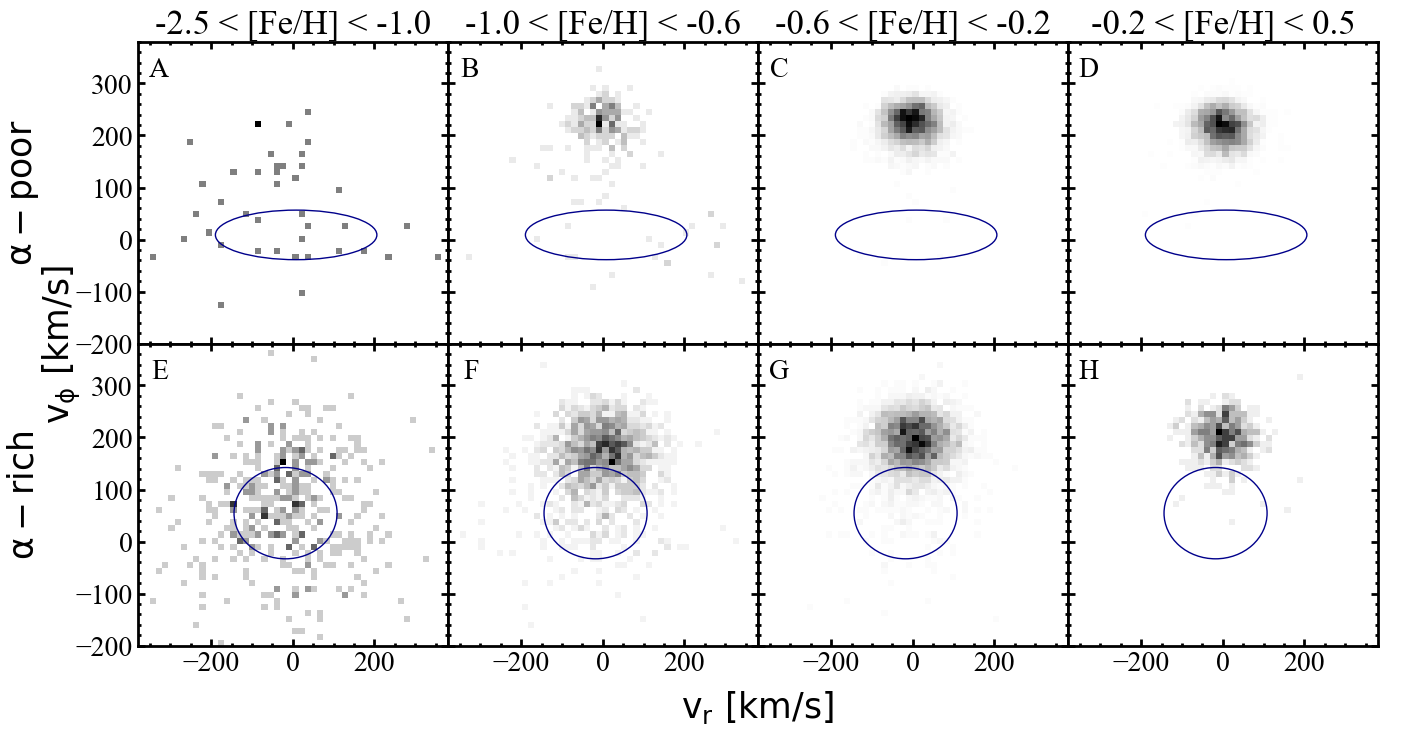}
\caption{Phase-space of the observed Gaia-ESO sample: Azimuthal velocity, $\vphi$, versus radial velocity, $\vr$, coloured in density. The upper panels represent the $\alpha$-poor population, and the lower panels represent the $\alpha$-rich population. Each population is split into two [Fe/H] bins, metal poor and metal solar-rich. The blue ellipse represents the halo distribution following the multi-Gaussian approach described in \cite{Belokurov2018} to determine halo contamination within the disc.}
\label{fig:Vr_2dhist}
\end{figure*}

Figure \ref{fig:Vr_2dhist} shows the distribution of our $\alpha$-poor and $\alpha$-rich samples in the plane of radial velocity, $V_{\rm r}$, versus azimuthal velocity, $V_{\phi}$ for different metallicity bins. The kinematic quantities were calculated using the positions, proper motions, and parallaxes from \textit{Gaia} DR3 and the \textit{astropy} package.

It is clearly seen that the majority of stars from both $\alpha$-rich and $\alpha$-poor populations are centred with $\vr\ \simeq 0$ kms$^{-1}$ and $\vphi\ \simeq 220$ kms$^{-1}$, consistent with the expectations for the Galactic disc \citep[][]{Ruchti2011, Navarro2011, Bensby2014}. With decreasing metallicity, the velocity dispersion in the radial direction increases, the mean rotation of stars decreases, and a counter-rotating component appears at [Fe/H] $\approx -0.6$, which is in line with previous studies of the disc \citep{Chiba2000,Fuhrmann2004,Deason2017}. Perhaps the most interesting feature of the observed distributions is the presence of a significant fraction of metal-poor, $-1 \lesssim$ [Fe/H] $\lesssim -0.6$, and $\alpha$-poor stars on disc orbits. In terms of kinematics, these stars are identical to the more metal-rich $\alpha-$poor population suggesting their thin disc origin. In Sect. \ref{sec:orbits}, we will perform a more detailed analysis of this group in terms of their integrals of motion in order to understand their properties. In the most metal-poor bin, [Fe/H] $\lesssim -1$ dex, low-$V_{\phi}$ stars with large radial velocities $| \vr | \gtrsim 200$ km/s appear. This population resembles the low-$\alpha$ metal-poor accreted halo component, which was identified in smaller targeted samples \citep[e.g.][]{Nissen2010}, and subsequently confirmed with other datasets \citep{Bergemann2017, Hayes_2018Sausage, Haywood_2018HaloDR2}. \citet{Belokurov2018} identified these stars as a population associated with a massive merger event around $8$ to $11$ Gyr ago, which was subsequently coined as the Gaia-Sausage \citep{Myeong2018} or the Gaia-Enceladus event \citep{Helmi2018}.

Combining \textit{Gaia} DR2 with APOGEE, \citet{Dimatteo2019} found that the accreted halo component is characterised by $\vphi$ $\approx$ 0 kms$^{-1}$ and is approximately Gaussian distributed in $\vr$, with the corresponding velocity dispersion of $\approx$ 120 kms$^{-1}$ \citep[][]{Lancaster2019}.  For highly prograde velocities,$\vphi$ > 100 kms$^{-1}$ \citep[][their Fig. 10]{Dimatteo2019}, however, the contribution of the accreted halo population is of the order of a few percent. We estimate the contamination by the halo stars in our sample following the multi-Gaussian decomposition approach by \citet{Belokurov2018}. The model estimates are performed separately for the $\alpha$-poor and the $\alpha$-rich populations. In short, we select all stars with $\vphi < 0$ and any $\vr$ value and model the bivariate distribution of $\vphi$ and $\vr$ by a Gaussian that is centred on $\vphi=0$. This resulting bivariate Gaussian function is assumed to represent the halo component of the entire stellar sample. Then, the resulting contamination fraction is calculated as the ratio of the number of stars in this Gaussian to to the total number of stars above a given $\vphi$ value. In Table \ref{tab:table1}, we show the resulting expected fractions of the contamination of our sample by halo stars, as predicted by our model for both $\alpha$ populations. Similar to \citet{Dimatteo2019}, we find the lowest value of this range to be $\sim$ 110 kms$^{-1}$ (Fig. \ref{fig:Vr_2dhist}). For $\vphi$ above this limit, the contamination by the halo is expected to be at the level of $<1 \%$ for the $\alpha$-poor population, as long as [Fe/H] $\gtrsim -1$. For the most metal-poor bin, [Fe/H] $\lesssim -1$, the contamination is $\sim 12 \%$. In the $\alpha$-rich population, the fractions are not too different in the metallicity bins [Fe/H] $\gtrsim -1$. However, as expected, the halo component becomes dominant over the disc for the most metal-poor $\alpha$-rich bin.
\begin{table}[h!]
\begin{center}
\caption{Contamination of the observed sample by halo stars in \%. See text.}
\label{tab:table1}
\begin{tabular}{l|c|c} 
\hline
\hline
Disc population & $\vphi \geq 110 \kms$ & $\vphi \geq 180 \kms$ \\
 & \% & \% \\
\hline
$\alpha$-poor & & \\
 $-2.5$ < \feh < $-$1.0  &  3.5 & 0.1 \\
 $-1.0$ < \feh < $-$0.6  &  0.2 & 0.0 \\
 $-0.6$ < \feh < $-$0.2  &  0.0 & 0.0 \\
 $-0.2$ < \feh < ~~0.5   &  0.0 & 0.0 \\
$\alpha$-rich & & \\
 $-2.5$ < \feh < $-$1.0  & 68.9 & 58.3 \\
 $-1.0$ < \feh < $-$0.6  &  3.3 & 1.9  \\
 $-0.6$ < \feh < $-$0.2  &  0.7 & 0.3  \\
 $-0.2$ < \feh < ~~0.5   &  0.0 & 0.0  \\
\hline
\hline
\end{tabular}
\end{center}
\end{table}
In addition, we compute the halo contamination through a slightly different model independent procedure described in detail in Appendix \ref{app:halo_cont}, where instead of fitting the $\vphi$ distribution, we calculate the contamination at $\vphi$ velocities in individual [Fe/H] bins by reflecting the $\vphi$ distribution across $\vphi=0$. The results of this calculation (Fig. \ref{fig:cont_running_ave}) confirm the decomposition based on \citet{Belokurov2018}, suggesting that above [Fe/H] $\gtrsim -1$ the observed stellar sample is primarily represented by stars with disc-like kinematics, and the contribution of accreted halo stars is marginal \citep[see also][]{Ruchti2011}. It is therefore safe to assume that stars above this metallicity with $\vphi \gtrsim 110$ kms$^{-1}$ are representative of the Galactic disc. Consequently, we rely on this criterion to define the disc sample and use it, in combination with the criteria to separate $\alpha$-poor and $\alpha$-rich populations (Sect. \ref{sec:chem_abund}), to investigate its evolutionary properties.
\subsection{Orbital Characteristics}\label{sec:orbits}
In order to further understand the dynamical properties of our chemically distinguished samples and to better characterise the old $\alpha$- and metal-poor sample, we take a step further and look at their action distribution. In principle, actions and their corresponding angles are just another set of canonical coordinates. However, for non-resonant orbits in axisymmetric potentials the actions are defined as the constants of motion, invariant under adiabatic changes and even mostly invariant under radial migration. Furthermore, in those potentials, the three conserved actions $J_\phi$, $J_r$ and $J_z$ correspond to our directions in cylindrical coordinates and have an intuitive meaning: $J_\phi$ is equal to specific angular momentum $L_z$; $J_R$ and $J_z$ are a measure of the radial, respectively vertical, excursion of the orbit around its guiding centre radius $R_g$ and the galactic plane. Using actions hence, by definition, allows us to fully classify any orbit by just three parameters. To obtain actions for our stellar sample, we used the \texttt{Agama} \citep{Agama2019} code with the Galactic potential from \citet{McMillan2017Pot}. 
In order to minimise the noise from distance errors, we introduce a parallax cut of $10 \%$ that reduces our full chemo-dynamical sample to $11\,137$ stars.
\begin{figure}[ht]
\includegraphics[width = \columnwidth]{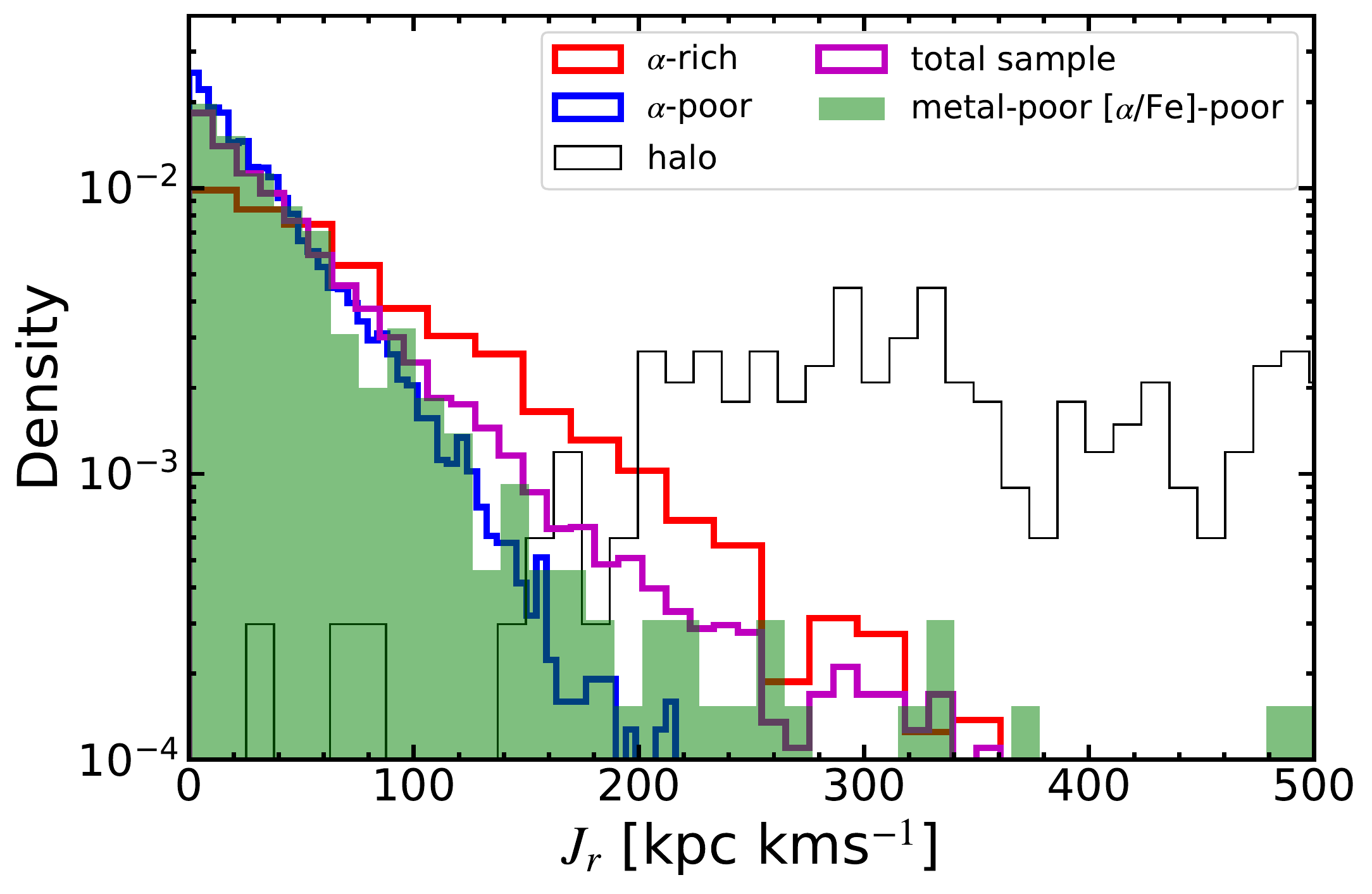}
\includegraphics[width = \columnwidth]{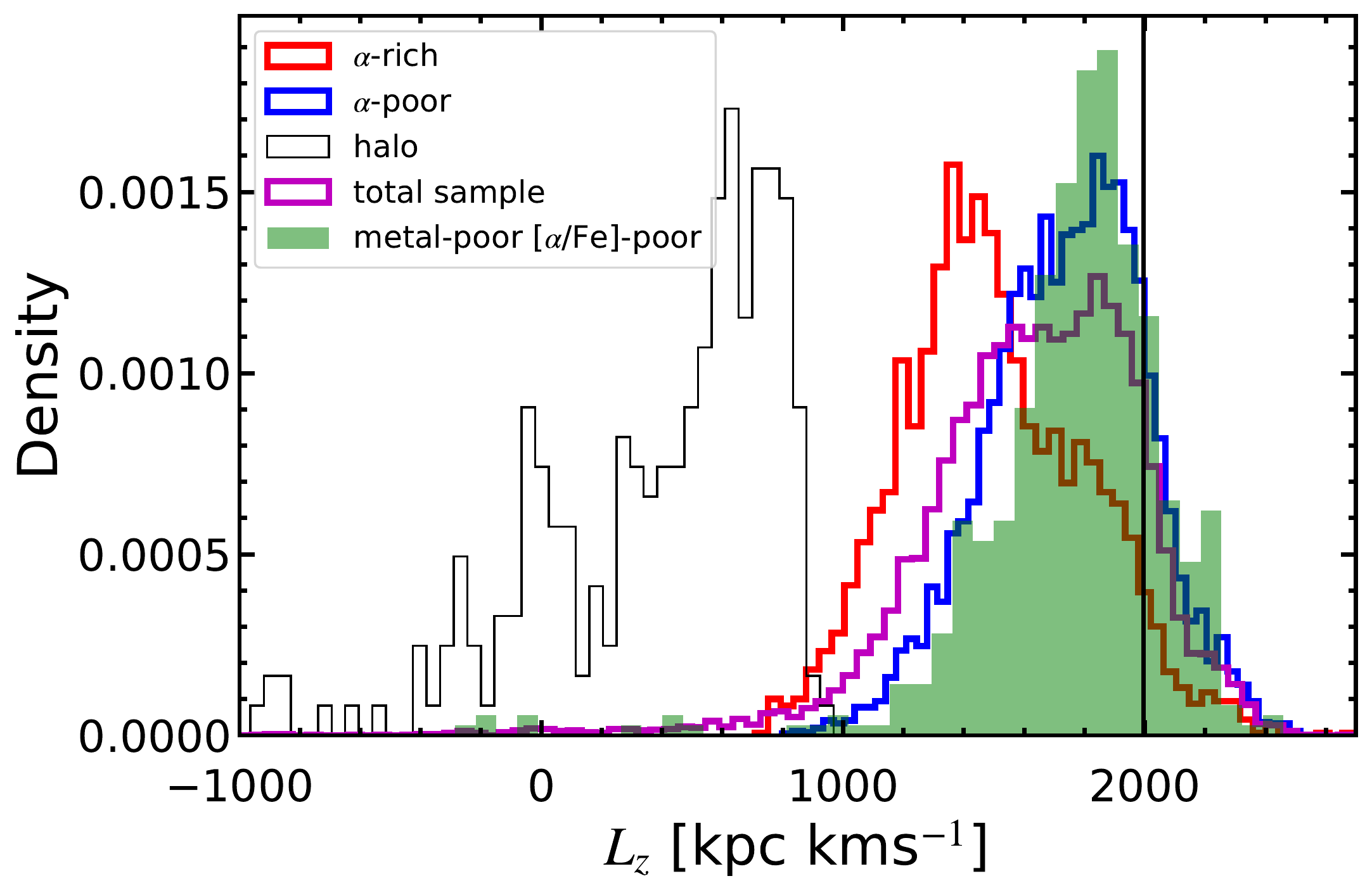}
\caption{Density distribution in actions \jr\ and \lz\ for our identified components. The black line in the lower panel represents the Solar \lz value of $1\,995$ kpc $\kms$.}
\label{fig:Jrexponential}
\end{figure}

Figure ~\ref{fig:Jrexponential} shows the resulting action distributions in \jr~and \lz, where we distinguish between the halo population, as identified kinematically in Sect. \ref{subsec:kinematics}, as well as the $\alpha$-poor and the $\alpha$-rich disc components defined chemically (see Sect. \ref{sec:chem_abund}). We also overplot the entire chemo-dynamical sample, and additionally highlight the other interesting sub-group: the metal-poor $\alpha$-poor sample, i.e. stars with [Fe/H] $< -0.5$ and [$\alpha$/Fe] $< +0.13$.
We find that both disc populations have their maximum value around zero value of \jr~ and that they are roughly exponentially distributed as expected by a pseudo-isothermal distribution function \citep{Carlberg1985,Binney2010DF}. They do show different exponential factors though, with the $\alpha$-poor disc characterised by a larger, more negative factor corresponding to the mean radial action in the sample $\hat{J_R}$.  This lower spread in \jr~ implies that the $\alpha$-poor stars are on radially less eccentric orbits than the $\alpha$-rich ones, i.e. the $\alpha$-poor sample is kinematically colder and on more circular orbits. Which is the reason for the often used identification between the chemically defined alpha-poor disc and the kinematically defined thin disc. The disc stars in our sample have |\lz| values that tend to be slightly lower compared to the solar value at about $2000$ kpc $\kms$. The $\alpha$-rich disc population shows somewhat lower |\lz|, i.e., it is dominated by stars on orbits with guiding centre radius further in the inner disk. This is a result of the shorter scale length of the $\alpha$-rich with respect to the $\alpha$-poor disc \citep{Bensby2011, Cheng2012ThickDisk}.

The $\alpha$-poor disc in Fig. \ref{fig:Jrexponential} has a very narrow distribution around \lz $\approx 1800$ kpc $\kms$ and it is skewed towards the larger values. The \lz~distribution of the $\alpha$-rich disc peaks at $\approx 1300$ to $1400$ kpc $\kms$ with a larger spread than the $\alpha$-poor disc. Similar distributions were reported by \citet{Bland-Hawthorn2019} and \citet{Gandhi2019}. The latter analysis suggests that the $\alpha$-poor and the $\alpha$-rich sequences have \lz\ range in $\approx 1800$ to $2100$ kpc $\kms$ and $\approx 1600$ to $1750$ kpc $\kms$, respectively over a range of ages (their Fig. 6). The overall orbital properties of the disc are also close to those reported by \citet[][]{Trick2019}, $\sim 1760$ kpc $\kms$, and to \citet{Buder2021}. \cite{Trick2019} used Gaia data, whereas our data based on the VLT spectra, and hence have more stars with lower R$_g$ and hence has a limited |\lz| range. The study \cite{Buder2021} does not distinguish between $\alpha$-rich and $\alpha$-poor in the kinematic analysis, the \lz~distribution of our total disc sample peaks at $1800$ kpc $\kms$ with an average value at $\approx 1700$ kpc $\kms$, which is consistent with their findings of the dominant fraction of stars with radial actions similar to the solar value. We note, however, that the GALAH sample is significantly brighter ($9 \gtrsim$ V $\gtrsim 14$), and hence more local compared to the Gaia-ESO. In addition (Sect. \ref{sec:observations}), the Gaia-ESO sample - owing to its photometric selection function - is somewhat biased against young stars with ages $< 7$ Gyr, which make a major fraction of the Galactic disc. One would therefore expect a better completeness of the thin disc in the GALAH distribution.

One particularly interesting feature of our sample is the presence of the $\alpha$-poor and metal-poor disc component. This component follows quite closely the \jr~and \lz~distribution of the $\alpha$-poor disc. Furthermore, with the exception of $10$ stars, all of these stars ($524$ stars with parallax-over-parallax-error cut of 10) are neither in the \lz~ nor at the high \jr~region that could be considered the Galactic halo, and their distribution is also not representative of that of the $\alpha$-rich disc. We carefully checked the observed data, as well as the stellar parameters and kinematic quantities, but did not find any evidence for a systematic error that could possibly lead to a mis-classification in terms of the population membership. The only distinct feature of this population is that it is concentrated at slightly higher |\lz|~values compared to the thin disc sample, which could possibly indicate that those are stars that formed in the outer disc and have since migrated inwards. However, currently we do not have any other means to provide a robust test of their origins.

Interestingly, evidence for a kinematically cold $\alpha$-poor and metal-poor disc population can also be seen in the analysis of the GALAH survey data by \citet{Bland-Hawthorn2019}. Their distributions suggest that for stars with [Fe/H] $< -0.5$ dex, the average \lz $\sim$is slightly higher than that of the average disc, closer to $1950$ kpc $\kms$, which is in good agreement with our results. \citet{xiang2022} find a non-negligible fraction of $\alpha$-poor metal-poor stars with \lz > 1500 kpc \kms\footnote{These studies use a left-Handed Galactocentric system which has the azimuthal velocity pointing in the direction of rotation and so, to compare, our subsequent \lz values are converted to our frame (RHS) by just flipping the sign \lz$_{RHS}$ = -\lz$_{LHS}$. Thus being consistent with our presented azimuthal velocities.}) and ages from 5 to 9 Gyr. This study came to a similar conclusion that this component may arise from stars born further out and which subsequently migrated inwards.  The origin of metal-poor $\alpha$-poor stars in the outer disc was also suggested by \citet{Haywood2013}, \citet{Buder2019}, and \citet{Wu2021}.
\begin{figure}
\centering
\includegraphics[width=0.9\columnwidth]{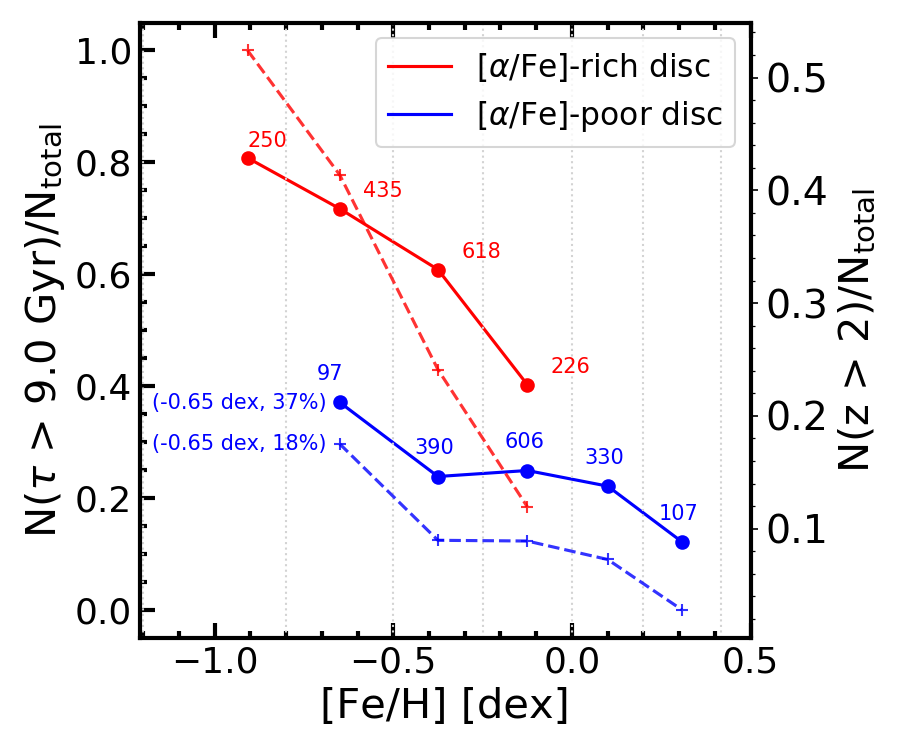}
\caption{Vertical axis is the Number of stars older than 9 Gyrs with respect to the total number (annotated as text) of $\alpha$-poor,rich stars (respectively) in a given [Fe/H] bin. Grey dotted lines represent [Fe/H] bin boundaries. Blue data represents $\alpha$-poor stars, red data represents $\alpha$-rich stars. The solid lines represent the left y-scale: fraction of stars with ages ($\tau$) more than 9 Gyr, the dashed lines represent the right y-scale: fraction of stars with redshift (z) more than 2.}
\label{fig:nstars9gyr}
\end{figure}
\subsection{Ages and abundances of disc stars}\label{agesabkin}
Finally, we explore the distribution of the two [$\alpha$/Fe] components with respect to stellar age. This analysis is limited to the final sample in order to avoid spurious trends caused by imprecise stellar parameters. Because of the very strong impact of the survey selection function on the completeness and statistical properties of younger populations (see Sect. \ref{sec:selection}), we do not explore the entire age distributions. Instead, we limit the analysis to the relative behaviour of old stellar components in the $\alpha$-poor and $\alpha$-rich sample. This choice is motivated by the fact that our samples appear to be rather complete only for ages in excess of $\approx 9$ Gyr. Therefore, this limit is used in Fig. \ref{fig:nstars9gyr} to show the fraction of very old stars in each $\alpha$ component to the total number of stars in the respective population against metallicity. In addition to the analysis of ages, we derive the relative fraction of stars with redshift (z) in excess of $\sim$ 2. This is driven by the peak epoch of star formation across the universe \citep{Madau2014}, to quantify the fraction of stars in both $\alpha$ sequences to investigate early formation of the Milky Way disc. 

To derive the redshift, ages were converted to z using the python module \textit{astropy.cosmology}. The cosmology chosen is the Flat Lambda-CDM model based Planck satellite data \citep{PlanckCollaboration2016} with a hubble constant of H$_0$ = 67.7 $\kms$Mpc$^{-1}$ and a fraction of observed density to critical density of $\Omega$ = 0.307. We chose this to be consistent with IllustrisTNG simulations \citep{Nelson2019}.

It is interesting that both components of the Galactic disc - and especially the $\alpha$-poor component - show a significant fraction of old stars at any [Fe/H]. As seen in Figure \ref{fig:nstars9gyr}, the fraction levels at $\approx 20\%$ at solar and slightly sub-solar metallicities, but it grows with decreasing [Fe/H] peaking at 37$\%$ for $\tau\ >\ 9$ Gyr and 18$\%$ for z $>\ 2$. For the $\alpha$-rich disc. The fraction of old stars attain their maximum of over $70 \%$ with $\tau\ >\ 9$ Gyr and over $40 \%$ with z $>\ 2$ at [Fe/H] $\lesssim -0.7$ dex. Investigating R-z, x-y, \lz-E$_{\rm tot}$ and other kinematic projections, we find that these old stars are unremarkable in the sense that statistically they have properties similar to the rest of the sample. The older stars are located more in the inner disc, exhibit disc like kinematics, and show no preference regarding to vertical altitude relative to the mid-plane.

At any given [Fe/H], the $\alpha$-rich population hosts a larger fraction of old stars compared to the $\alpha$-poor population. Both components furthermore span a range of ages. Most stars in the $\alpha$-rich population have ages of $\sim 8$ to $\sim 12$ Gyrs, in line with the results in the literature  \citep[e.g.][]{Bensby2005,Haywood2013,Hayden2017,Buder2019,xiang2022}. In contrast, the $\alpha$-poor population is characterised by a much wider distribution of ages from a few to $\sim 10$ Gyr. We stress that especially for metal-poor stars, age estimates appear to be most reliable (see the discussion in Sect. \ref{sec:stellarparameters}), also the uncertainty of [Mg/Fe]-ratios is expected to be about $0.06$ dex, modulo the accuracy that is set by the physical models and would possibly imply an additional systematic error component of a similar order of magnitude \citep[][]{Bergemann2017}, but the latter would affect any sample not analysed with full 3D NLTE models. It is, therefore, rather unlikely that this peculiar group represents mis-classified stars. Before we proceed with the interpretation, we remind the reader that our distributions are likely heavily biased against young stars, hence the fraction of old stars at metallicities close to solar are likely significantly over-estimated.

The presence of old $\alpha$-poor stars is intriguing although not fully unexpected. Among the earlier studies, \citet{Haywood2013} and \citet{Hayden2017} remarked on the significant temporal and chemical ([Fe/H]) overlap of the $\alpha$-rich and $\alpha$-poor components of the Galactic disc, studying the properties of TO and subgiant stars in the AMBRE:HARPS survey. \citet{silvaaguirre18} identified a population of old $\alpha$-poor stars through asteroseismic age dating of APOGEE targets in the Kepler field. They report an overlap in age from 8 - 14 Gyrs between the $\alpha$-poor and $\alpha$-rich components. Using APOGEE and SEGUE data, \citet{laporte20} also report entirely old low-$\alpha$ stellar populations with ages between 12 and 8 Gyr in the Anticenter Stream, which shows a dearth of younger stars in its cumulative age distribution when compared to the Monoceros Ring \citep{newberg02} suggesting an early decoupling from the rest of the Galactic disc following an interaction with a satellite \citep[e.g.][]{laporte19a,naidu21}. Some fraction of $\alpha$-poor with ages up to $9-10$ Gyr is also evident in the results by \citet{xiang2022} 
Similar distributions are seen in the results based on the APOGEE data by \citet{Ciuca2021} and \citet{beraldoesilva21}, who find $\alpha$-poor stars with ages up to $\sim 12$ Gyr spanning the entire range of metallicity, $-0.7 \lesssim \feh \lesssim +0.4$, in the solar neighbourhood. The study by \citet{beraldoesilva21} is particularly relevant in the context of our work, as the spatial coverage is similar, R$_{\rm helio}$ $\sim 2$ kpc, and the targeted stellar population (subgiants and the lower part of the RGB branch) overlaps with the properties of our observed Gaia-ESO sample. \citet{Feuillet2019} find rather tight age-$\afe$ relationships, with some evidence for the presence of old $\alpha$-poor stars.

In the next section, we will explore the potential implications of the temporal overlap of the $\alpha$-poor and $\alpha$-rich disc populations within the context of Galaxy formation.
\section{Discussion}
There are many debated scenarios for the formation of $\alpha$-rich and $\alpha$-poor sequences. Among them is the canonical two-infall model with a star-formation hiatus \citep[e.g.][]{Chiappini1997}, also extended to a three-infall model in \citet{Spitoni2019}.
Another scenario is formation of the $\alpha$-rich component from the primordial thin disc, i.e the high-redshift clumpy star formation \citep[e.g.][]{Bournaud2009,Clarke2019}. A scenario, primarily realised in cosmological simulations of galaxy formation, is associated with mergers, accretion of metal-poor gas, and/or the differential evolution of the inner and outer discs \citep[e.g.][]{Brook2012,Grand2018,Buck2020,Agertz2021}.
The study of \citet{Haywood2019} presents two evolutionary pathways with respect to the inner and outer disc as a combination of a pre-enrichment provided by the thick disc, which formed from a turbulent gaseous disc, a quenching phase, and a sudden dilution episode by more H-rich gas that results in a metal-poor $\alpha$-sequence.

Our results suggest that the growth of the $\alpha$-poor and $\alpha$-rich disc components accompanied each other at least during a period of a several Gyr, challenging the strictly sequential scenarios, in which thin discs form after much of the thick disc is in place. While the $\alpha$-rich disc component is on average older and slightly more metal-poor than the $\alpha$-poor component, we also find  many $\alpha$-poor stars that are as old as those in the $\alpha$-rich population, that is $9$ Gyr and older, and even pre-date the Gaia-Sausage-Enceladus merger \citep{Belokurov2018, Helmi2018}. Thus, this major merger event does not seem to be the trigger of the low-$\alpha$ metal-poor disc formation. In the cosmological simulations, multiple formation pathways are realised. For example, the thin disc (or $\alpha$-poor sequence) may develop following minor gas-rich mergers, which in simulations by \citet[][]{Buck2020} happens at the look-back times of $\sim 7-9$ Gyr. The NIHAO-UHD simulations also exhibit metal-poor, $\alpha$-poor and old stars - as old as $\sim$11 Gyr \citep[][their Fig.3]{Buck2020}. These stars originate from early accreted dwarf galaxies onto the proto-Milky Way and happen to reside in the disc at the solar radius at present day. However, it has not been studied if these stars are on disc like orbits, nor if their relative fraction matches the 20-30\% of metal-poor, $\alpha$-poor and old stars as observed in this study. 
In the VINTERGATAN I simulations \citep{Agertz2021}, the spatially extended $\alpha$-poor disc is formed very early, at $z \sim 1.5$, following a major merger event and thus can be seen as co-evolving with the inner old $\alpha$-rich disc population. 
Similarly, in \citet{Grand2018}, early gas-rich mergers may trigger the more centrally-concentrated formation of the thick disc and so the thin disc component then subsequently grows in the inside-out fashion via following the phase of the gas disc contraction (lowering the star formation rate) and metal-poor accretion. Also in this study, the growth is associated with $\tau \lesssim 8$ Gyr. In the simulations of \citet{Brook2012}, a similar pathway is seen, with the thick disc forming earlier primarily from the gas assembled through high-redshift gas-rich mergers, and the $\alpha$-poor sequence starting to form at $\tau \sim 6-7$ Gyr from smoothly and continuously accreting gas. This formation sequence bears resemblance to the scenario discussed on the basis of VINTERGATAN II simulations in \citet{Renaud2021}. At this stage, it seems that it is rather uncommon for the galaxy formation simulations to initiate a very early formation of the thin ($\alpha$-poor) disc, and thus, it is not yet clear whether these models explain our findings.

Another relevant scenario is the clumpy star formation, that can be understood as thick discs forming out of primordial thin discs \citep[e.g.][]{Bournaud2009,Agertz2009, Clarke2019}. Gas-rich fragments that develop in early turbulent  discs are characterised by high star formation rate density, effectively self-enriching to a high-$\alpha$ population. In this scenario, both $\alpha$-rich and $\alpha$-poor sequences start forming very early, and share a limited period of co-evolution at ages $\sim 9-10$ Gyr \citep[][their Fig. 11]{Clarke2019}, before star formation in the $\alpha$-rich clumps is halted at $\sim 6$ Gyr, whereas the $\alpha$-poor sequence continues forming with a low star formation rate to present day. Hence, our results can be well understood in the framework of the clumpy disc formation model.

Finally, in the scenario proposed by \citet{Haywood2019}, the origin of metal-poor low-$\alpha$ stars is associated with the outer disc and in this scenario, $\alpha$-poor stars as old as $10$ Gyrs emerge. We note, however, that this scenario has been primarily explored within the framework of a closed box model with arbitrary parameters. For example, this model relies on a semi-empirical star formation history \citep{Snaith2015}, which, in turn, is constructed by fitting the observations, e.g. the observed age-[Si/Fe] distributions for stars in the solar neighbourhood adopted from \citet{Adibekyan2012}, for which ages were derived by \citet{Haywood2013}. That stellar sample is bright (V $\lesssim 11$) and it was originally developed for a different purpose, namely the HARPS GTO planet search program \citep{Mayor2003, Santos2011}, and its completeness in terms of Galactic structure has never been demonstrated. Thus, the model proposed by \citet{Haywood2019} cannot be used to directly (and independently) to interpret our observational results, as our sample and its spatial, kinematical and chemical distributions are completely different from what the HARPS planet-search observational sample represents.
\section{Conclusions}\label{sec:conclusion}
In this work, we study the chemical and kinematical properties of the Galactic disc using $13\,426$ stars observed by the Gaia-ESO survey \citep{Gilmore2012,Randich2013}. We used spectra from the fourth public data release (DR4) of the survey along with \textit{Gaia} EDR3 kinematics and photometry to derive stellar parameters and NLTE chemical abundances using the SAPP \citep{Gent2022a}. The majority of stars in our sample span the range of $6<R<10$ kpc in Galactocentric radius and $|Z|<2\,\rm{kpc}$ in vertical distance from the plane. We also determined ages for upper main-sequence and subgiant stars and explored the influence of the Gaia-ESO photometric selection function on the statistical properties of the sample. Through a comprehensive analysis of stellar parameters and ages using different methods and numerical codes, we show that the quality of results is sufficiently high to allow a quantitative analysis of astrophysical distributions, with a total uncertainty of abundances and ages of $\sim 10 \%$. 

Similar to previous studies, we find that our sample shows a prominent bi-modality in the [Mg/Fe] abundance ratios over a range of metallicities, [Fe/H] from $\sim -1$ to $\sim 0.0$ dex. Through the analysis of velocity distributions, we show that the contamination of the halo at [Fe/H] $\gtrsim -1$ dex is marginal and does not exceed $\sim 3 \%$. Our disc  population contains a significant fraction of metal-poor and $\alpha$-poor stars, [Fe/H] $\lesssim -0.5$ and [Mg/Fe] $\lesssim 0.05$, that are kinematically cold and are unlikely to represent an accreted halo population. We note that metal-poor $\alpha$-poor stars were also reported in other studies \citep[e.g.][]{Adibekyan2012}. 

The specific angular momentum distributions of the stellar populations in our sample are slightly different, confirming previous studies \citep[e.g.][]{Gandhi2019}. The $\alpha$-rich disc component has a rather extended  distribution $800 \lesssim$ \lz $\lesssim 2200$, and a maximum at \lz $\sim 1400$ kpc $\kms$. In comparison, the $\alpha$-poor disc has a narrower distribution of \lz~values, with the maximum at \lz $\sim 1800$ kpc $\kms$ and a dispersion of $\sim$ 260 kpc $\kms$. The metal-poor $\alpha$-poor sub-group has an even narrower \lz~dispersion and is skewed to larger \lz~values compared to more metal-rich stars, that is, larger guiding radii at a given circular velocity, which may indicate that these stars migrated from the outer disc inwards.

From the analysis of the age distributions, we find an interesting group of $\alpha$-poor disc stars that are old, with ages in excess of $\approx 9$ Gyrs. This population represents only a small fraction of the entire disc, however, compared to all disc stars formed at a lookback time of $\gtrsim 9$ Gyr, this group constitutes a non-negligible fraction of $\sim 20 \%$. The $\alpha$-rich disc extends out to ages over 12 Gyr, and the $\alpha$-poor component with $-0.7 \lesssim$ [Fe/H] $\lesssim 0.3$ is present already as early as 11 Gyr ago (at redshift z $\sim$ 2). The temporal extent of the sequences suggests that the $\alpha$-poor and $\alpha$-rich disc populations shared a limited period of co-evolution from $\sim$ 9 to 11 Gyr. Our study is not the first to report old $\alpha$-poor stars \citep[see e.g.][]{Haywood2013, Hayden2017,silvaaguirre18,Ciuca2021,laporte20,beraldoesilva21}, thus their existence appears to be on a firm footing.

The early co-evolution of the $\alpha$-rich and $\alpha$-poor populations can be explained by scenarios, in which the thin disc does not require a thick disc to be present (with or without metal-poor gas infall), before star formation in the $\alpha$-poor sequence is initiated. According to our data, the chemical thin disc shall be in place early, at look-back times corresponding to the redshift $z\sim2$. This result challenges the canonical picture of sequential disc formal in multi-infall models or the thin disc formation triggered by a gas-rich merger, as in both cases simulations predict thin disc stars not older than $\sim 8$ Gyr \citep{Grand2018,Buck2020}. At this stage, our results can be explained within the framework of the clumpy and distributed star formation scenario \citep{Bournaud2009, Clarke2019}.

Recent studies, such as \citet{Sestito2021} and \citet{Chen2023}, explore the origins of very metal-poor stars on disky orbits. Although they find many such stars in the models, associating them with ex-situ formation, the analysis is typically limited to very low metallicities. We therefore encourage a closer look at the formation of metal-poor $\alpha$-poor sequences in cosmological zoom-in simulations of galaxy formation, probing the entire [Fe/H] range, as e.g. was done in \citet{Sotillo-Ramos2023}.
Also on the observational side, a detailed investigation of the orbital and chemical properties of the early, possibly primordial, Galactic disc would be a very interesting venue for the future investigations, e.g with the surveys like 4MIDABLE-HR and 4MIDABLE-LR surveys \citep{Bensby2019, Bergemann2019, Chiappini2019} on 4MOST \citep{deJong2019}, which will provide a much more complete and homogeneous spatial and temporal coverage of the disc, possibly allowing to distinguish between the disc formation scenarios. 

\section*{Acknowledgements}
We thank Ralph Sch\"{o}nrich for valuable discussions.
This research made use of Astropy,\footnote{http://www.astropy.org} a community-developed core Python package for Astronomy \citep{astropy2013, astropy2018}. 
JF acknowledges support from University College London’s Graduate Research Scholarships and the MPIA visitor programme. MB is supported through the Lise Meitner grant from the Max Planck Society. We acknowledge support by the Collaborative Research centre SFB 881 (projects A5, A10), Heidelberg University, of the Deutsche Forschungsgemeinschaft (DFG, German Research Foundation). This project has received funding from the European Research Council (ERC) under the European Unions Horizon 2020 research and innovation programme (Grant agreement No. 949173). AS acknowledges grants PID2019-108709GB-I00 from Ministry of Science and Innovation (MICINN, Spain), Spanish program Unidad de Excelencia Mar\'{i}a de Maeztu CEX2020-001058-M, 2021-SGR-1526 (Generalitat de Catalunya), and support from ChETEC-INFRA (EU project no. 101008324). CL acknowledges funding from the European Research Council (ERC) under the European Union’s Horizon 2020 research and innovation programme (grant agreement No. 852839). TB’s contribution to this project was made possible by funding from the Carl Zeiss Foundation. Based on data products from observations made with ESO Telescopes at the La Silla Paranal Observatory under programme ID 188.B-3002. These data products have been processed by the Cambridge Astronomy Survey Unit (CASU) at the Institute of Astronomy, University of Cambridge, and by the FLAMES/UVES reduction team at INAF/Osservatorio Astrofisico di Arcetri. These data have been obtained from the Gaia-ESO Survey Data Archive, prepared and hosted by the Wide Field Astronomy Unit, Institute for Astronomy, University of Edinburgh, which is funded by the UK Science and Technology Facilities Council. This work presents results from the European Space Agency (ESA) space mission Gaia. Gaia data are being processed by the Gaia Data Processing and Analysis Consortium (DPAC). Funding for the DPAC is provided by national institutions, in particular the institutions participating in the Gaia MultiLateral Agreement (MLA). The Gaia mission website is https://www.cosmos.esa.int/gaia. The Gaia archive website is https://archives.esac.esa.int/gaia.

 \bibliographystyle{aa}
 \bibliography{references} 

\appendix

\section{Halo contamination}
\label{app:halo_cont}

\begin{figure}
    \centering
    \includegraphics[width=0.8\columnwidth]{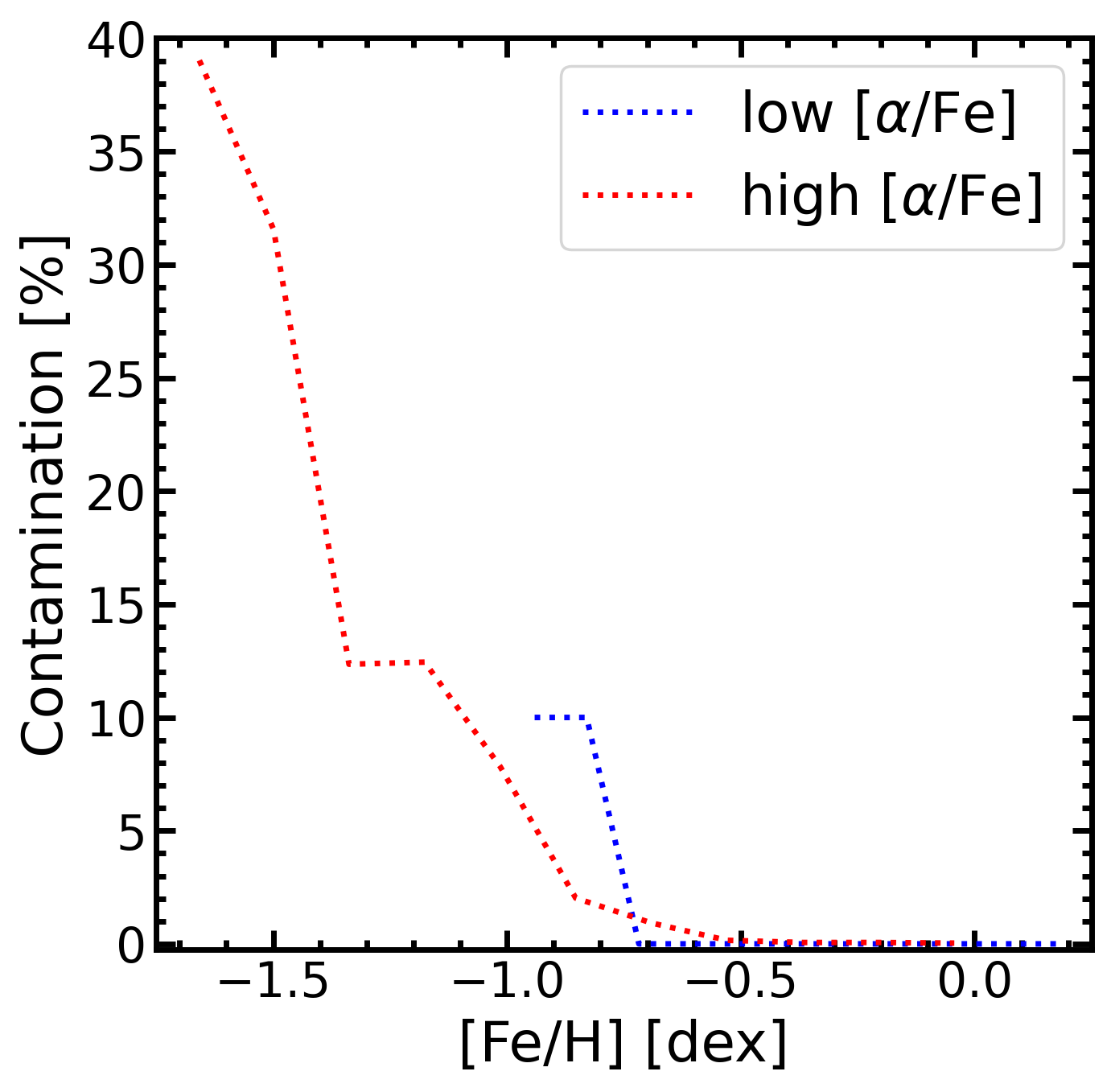}
    \caption{Running average of halo contamination in the disc in variable [Fe/H] bins with $\vphi$ > 110 Kms$^{-1}$. The red dotted line represents high $\alpha$ and the blue dotted line represents low $\alpha$.}
    \label{fig:cont_running_ave}
\end{figure}

Figure \ref{fig:cont_running_ave} shows the determination of halo contamination in the thick and thin disc for varying [Fe/H] bins as an alternative method. Similarly described in Sec. \ref{subsec:kinematics}, we analyse stars with $\vphi$ < 0 Kms$^{-1}$, assume symmetry in $\vphi$ distribution with respect to $\vphi=0$ and therefore determine the number of halo stars with positive azimuthal velocities. The contamination is determined by inspecting the number of stars with $\vphi$ > 110 Kms${-1}$ and comparing that to the number of stars in total above the velocity cut. This is determined for each bin of [Fe/H] and so a running average is calculated as opposed to fitting a velocity ellipsoid. This allows us to determine how contamination depends on metallicty and therefore informs the [Fe/H] limit for each alpha- population. Assuming a cut at [Fe/H] = -1, the average halo contamination is less than 10$\%$. 
 
\section{Age-Metallicity PDFs}

\begin{figure*}
\centering
\vbox{
\hbox{
\includegraphics[width=\columnwidth]{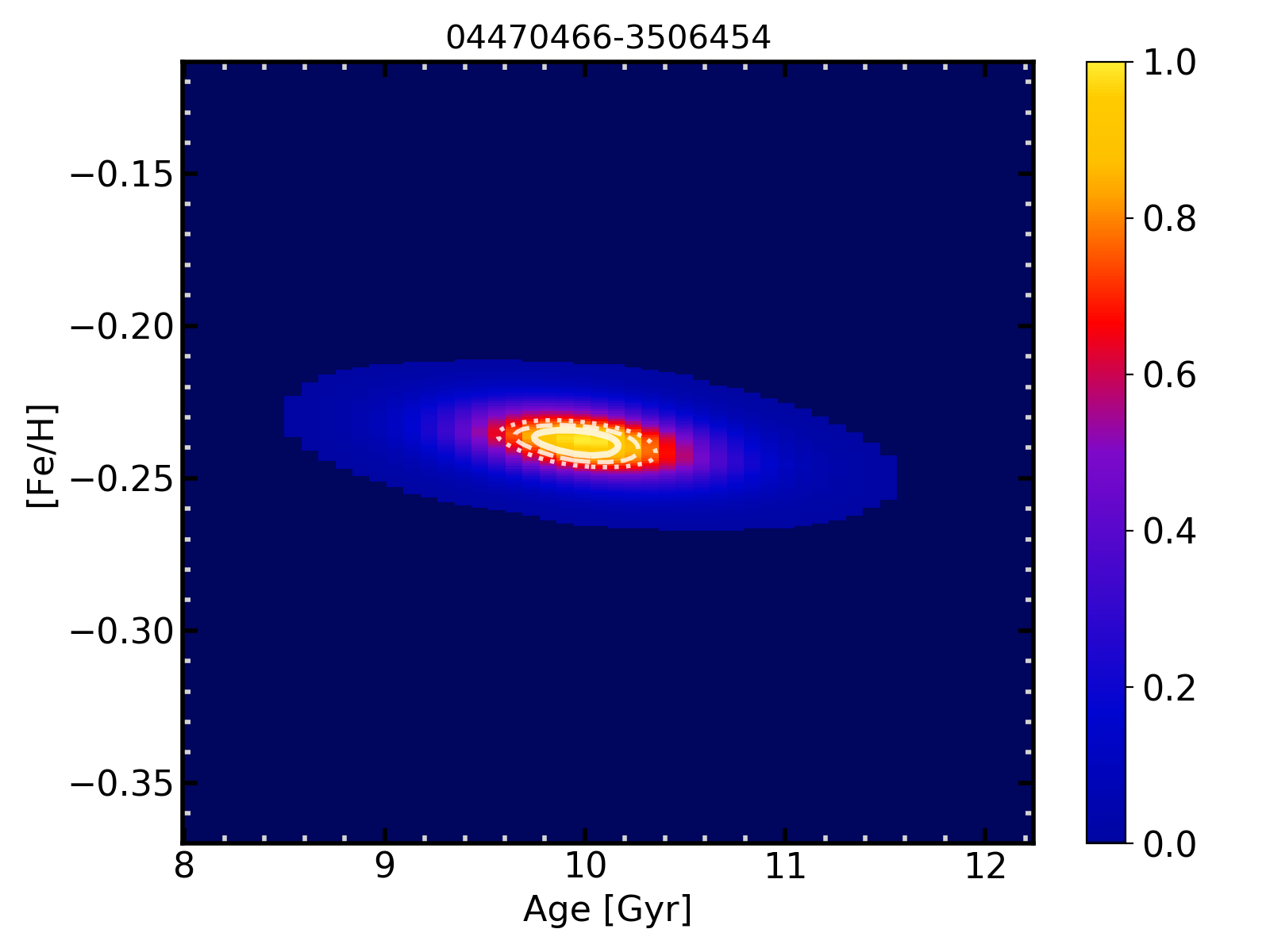}
\includegraphics[width=\columnwidth]{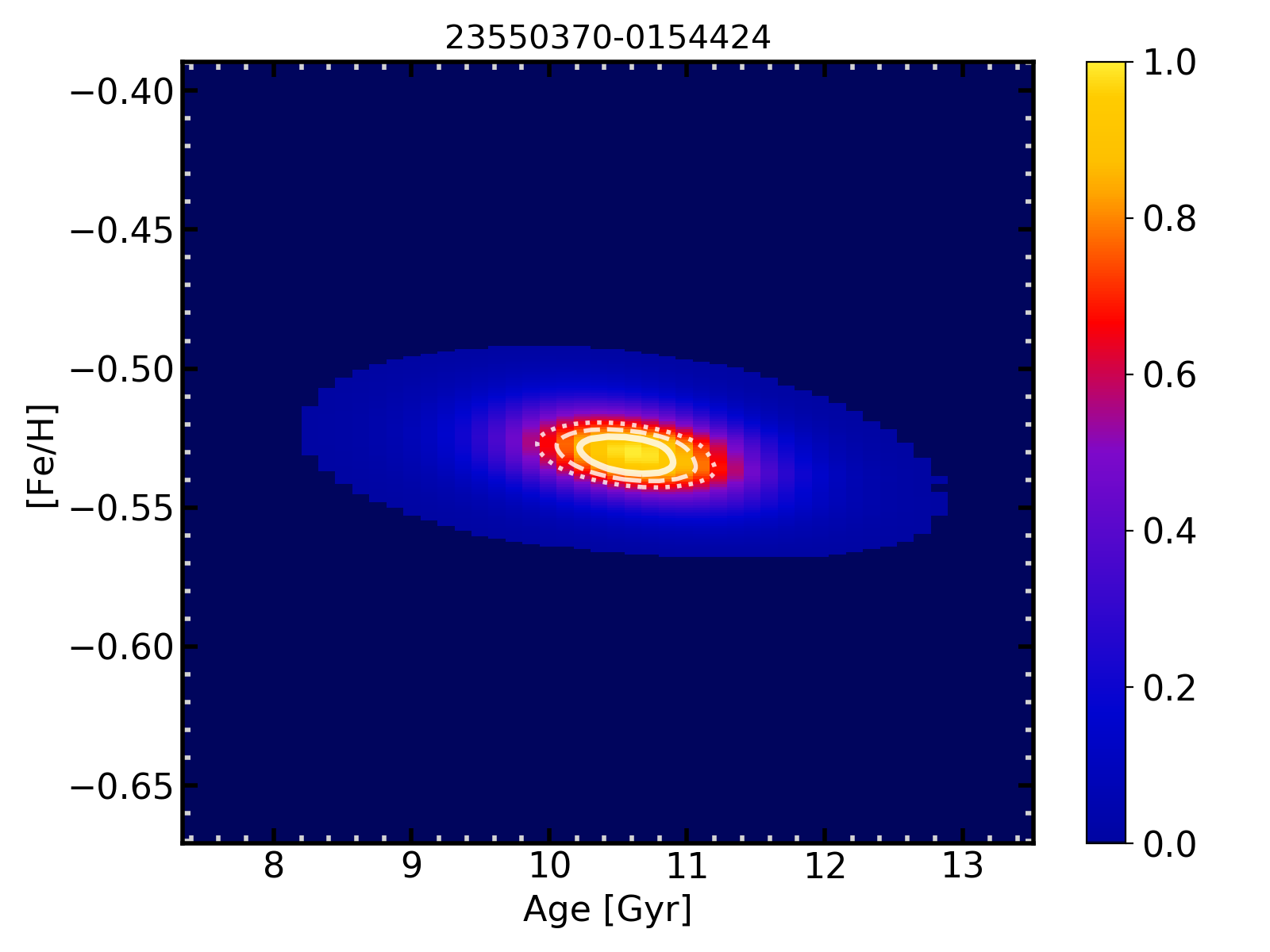}
}
\hbox{
\includegraphics[width=\columnwidth]{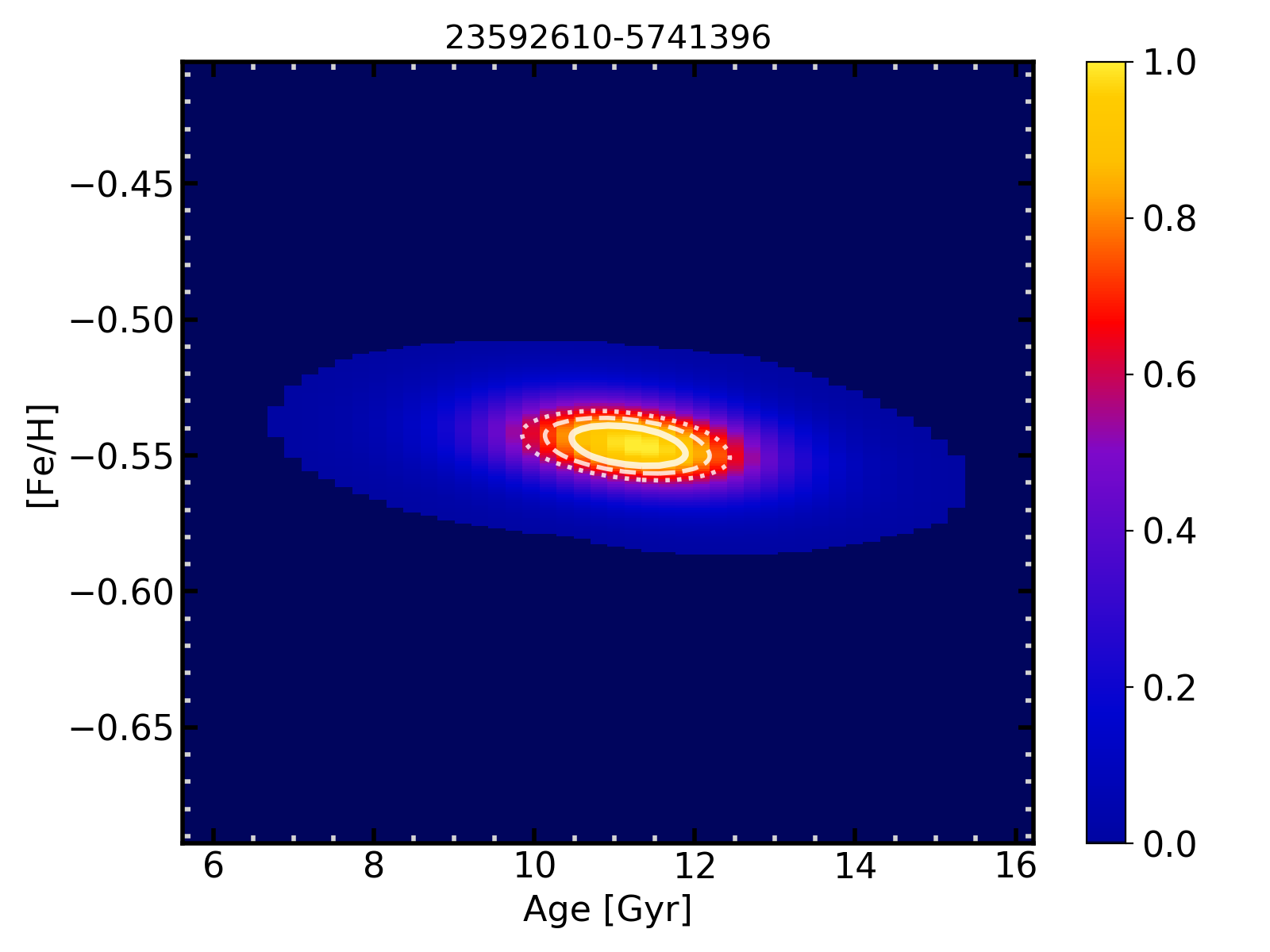}
\includegraphics[width=\columnwidth]{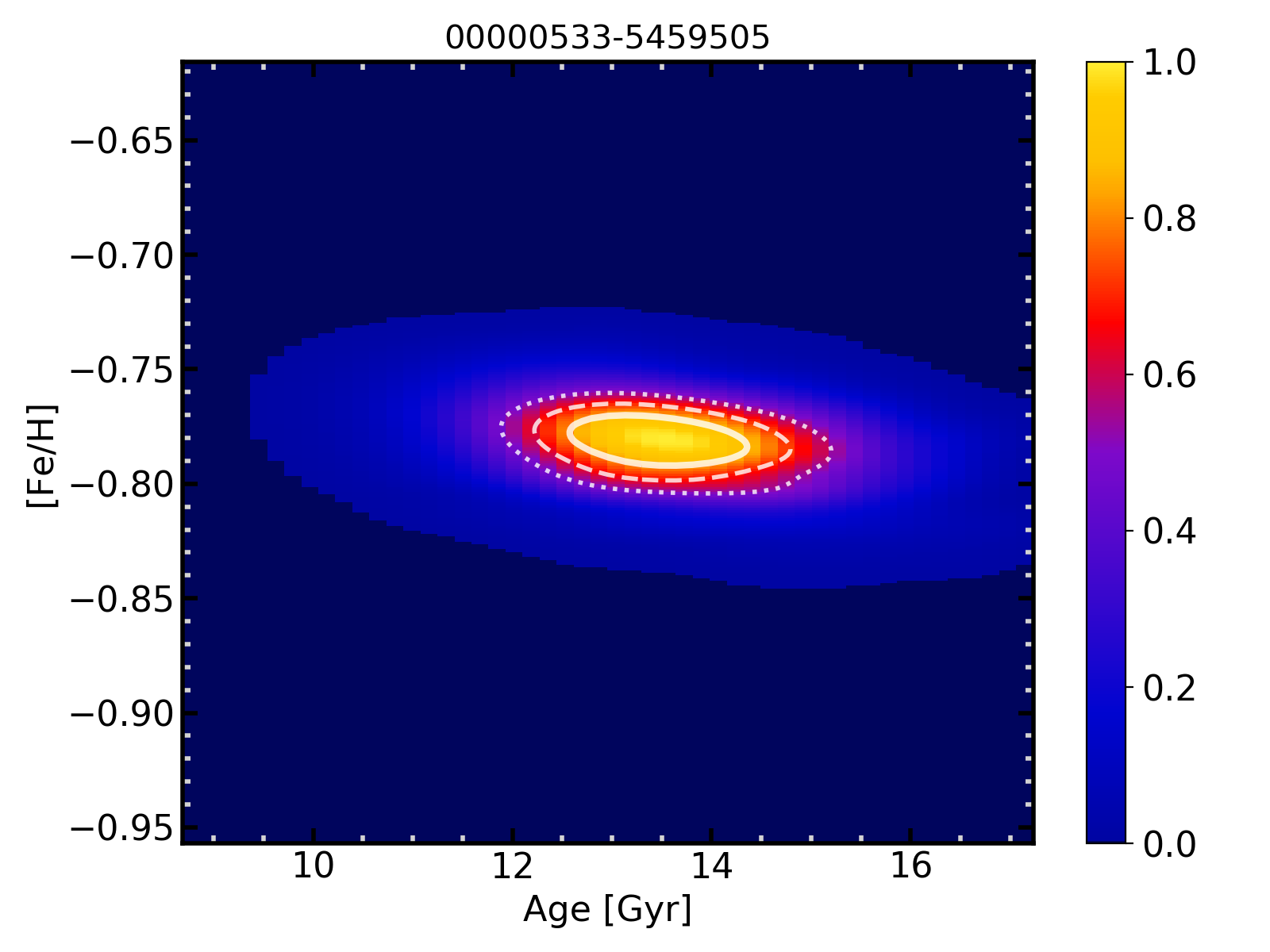}
}
}
\caption{Age-metallicity PDFs for 4 $\alpha$-poor and metal-poor stars. Each panel is a 2D histogram of the BeSPP posterior PDF in the [Fe/H]-Age projection with 1, 2, 3 -$\sigma$ contours. The colour scale is probability.}
\label{fig:AMR_pdf}
\end{figure*}

Figure \ref{fig:AMR_pdf} shows age-metallicity PDFs for 4 $\alpha$-poor and metal-poor stars. The internal ID of each star according to the Gaia-ESO catalogue is provided in the title.  The colour scale represents normalised probability density.

\end{document}